\documentclass[twocolumn]{aastex6}

\usepackage{graphicx,epstopdf,amsmath}
\usepackage{import}

\usepackage{color}

\newcommand{\phpl}{PhPl}

\newcommand{\jcoph}{JCoPh}
\newcommand{\cophc}{CoPhC}

\newcommand{\soph}{SoPh}

\newcommand{\natur}{Natur}

\newcommand{\gapfd}{GApFD}

\shorttitle{A Model for Coronal-Hole Bright Points and Jets due to MMEs}
\shortauthors{Wyper et al.}

\begin{document}

\title{A Model for Coronal Hole Bright Points and Jets \\ due to Moving Magnetic Elements}

\author{P.~F.~Wyper} 
\affil{Department of Mathematical Sciences, Durham University, Durham, DH1 3LE, UK}
\email{peter.f.wyper@durham.ac.uk}

\author{C.~R.~DeVore} 
\email{c.richard.devore@nasa.gov}

\author{J.~T.~Karpen} 
\email{judith.t.karpen@nasa.gov}

\author{S.~K.~Antiochos} 
\email{spiro.antiochos@nasa.gov}
\affil{Heliophysics Science Division, NASA Goddard Space Flight Center, 8800 Greenbelt Rd, Greenbelt, MD 20771, USA}

\author{A.~R.~Yeates} 
\affil{Department of Mathematical Sciences, Durham University, Durham, DH1 3LE, UK}
\email{anthony.yeates@durham.ac.uk}

\begin{abstract}
Coronal jets and bright points occur prolifically in predominantly unipolar magnetic regions, such as coronal holes, where they appear above minority-polarity intrusions. Intermittent low-level reconnection and explosive, high-energy-release reconnection above these intrusions are thought to generate bright points and jets, respectively. The magnetic field above the intrusions possesses a spine-fan topology with a coronal null point. The movement of magnetic flux by surface convection adds free energy to this field, forming current sheets and inducing reconnection. We conducted three-dimensional magnetohydrodynamic simulations of moving magnetic elements as a model for coronal jets and bright points. A single minority-polarity concentration was subjected to three different experiments: a large-scale surface flow that sheared part of the separatrix surface only; a large-scale surface flow that also sheared part of the polarity inversion line surrounding the minority flux; and the latter flow setup plus a ``fly-by'' of a majority-polarity concentration past the moving minority-polarity element. We found that different bright-point morphologies, from simple loops to sigmoids, were created. When only the field near the separatrix was sheared, steady interchange reconnection modulated by quasi-periodic, low-intensity bursts of reconnection occurred, suggestive of a bright point with periodically varying intensity. When the field near the PIL was strongly sheared, on the other hand, filament channels repeatedly formed and erupted via the breakout mechanism, explosively increasing the interchange reconnection and generating non-helical jets. The fly-by produced even more energetic and explosive jets. Our results explain several key aspects of coronal-hole bright points and jets, and the relationships between them.
\end{abstract}



\section{Introduction}
Small brightenings and impulsive flows are found throughout the ``quiet" solar corona, in association with photospheric magnetic-field concentrations and parasitic-polarity intrusions.  These features are most visible as bright points and jets in coronal holes (CHs), where the background magnetic field is largely unipolar and open and the ambient plasma is dark in EUV and X-ray wavelengths.  Coronal bright points are seen as enhanced EUV and X-ray emission from small regions with diameters on the order of 10-50 arcsec, and lifetimes of 3-60 hr in EUV \citep{Zhang2001,Mou2016} and up to $8$ hr in X-rays \citep{Golub1974}. A puzzling feature of most bright points is that their intensity varies periodically, with periods ranging from a few minutes up to a couple of hours \citep{Kariyappa2008,Tian2008}.  Their internal morphology appears to be similar in coronal holes and quiet Sun  \citep{Habbal1990,Galsgaard2017}: bright points can contain a sigmoid \citep[e.g.,][]{Brown2001}, a few parallel loops \citep{Zhang2012}, or an anemone \citep[e.g.,][]{Shibata1994}. In many cases, bright points are associated with moving magnetic elements (MMEs), sometimes with opposite-polarity concentrations separating from each other due to flux emergence, at other times converging toward each other to coalesce and cancel \citep{Webb1993,Mou2016}. These associations with interactions between opposite-polarity magnetic fields have led to broad acceptance that bright points ultimately derive their energy from magnetic reconnection and the ensuing acceleration and heating of the entrained plasma.

Long-lived bright points are frequently observed to produce coronal jets: impulsive, collimated flows of dense, hot plasma, which are launched low in the atmosphere and are guided along the ambient coronal magnetic field \citep{Shimojo1996,Nistico2009,Raouafi2016}. Jets have much shorter lifetimes (a few minutes) than bright points \citep{Savcheva2007}, so an individual bright point can produce multiple jets over its lifetime. Some jets extend so far into the corona that they can be observed in scattered white light in the inner heliosphere \citep[e.g.,][]{Wang1998}, thus contributing mass, momentum, and distinct structures to the solar wind. High-resolution observations have revealed that many, if not most, jets contain miniature filaments (cool plasma) and/or sigmoids (hot plasma) that erupt to generate the jet \citep[e.g.,][]{Innes2009,Raouafi2010,Shen2012,Hong2014,Hong2016,Sterling2015,Sterling2016,Kumar2018}. Typically, though not always, these jets have a strong helical flow component.  A relationship between bright points, jets, and the diffuse, persistent columns known as coronal plumes has long been suspected, but has proven difficult to verify and explain. Thus far the strongest connection appears to be the existence of tiny ``jetlets" observed within some plumes \citep{Raouafi2008,Raouafi2014}, but it is unclear whether all plumes are composed of many such impulsive events or whether a different mechanism (e.g., weak, quasi-steady null-point reconnection) is responsible for the enhanced density and flows of plumes. 

On close inspection, CH jet sources and anemone-type bright points \citep[e.g.,][]{Galsgaard2017} match the magnetic topology of an embedded bipole: a three-dimensional (3D) coronal magnetic null point, with associated inner and outer spine lines and a fan separatrix surface \citep{Antiochos1990,Lau1990}. The photospheric manifestation of this configuration is a minority-polarity intrusion within the majority-polarity magnetic field.  The fan surface separates the 
closed magnetic flux beneath the null point from the globally open flux above and away from the null point. Relative motions of the two flux systems can readily distort the null to form a current sheet there \citep{Antiochos1996}, setting the stage for interchange reconnection between open and closed field lines and associated plasma flows and heating.  It is broadly accepted that coronal jets are driven by the onset of explosive magnetic reconnection, while {some observational studies have speculated that} more gradual reconnection in the same magnetic structure could explain a long-duration bright point \citep[e.g.,][]{Doschek2010,Pucci2012,Zhang2012}. 

In previous work, we and our colleagues have investigated the generation of CH jets within the embedded-bipole model and its null-point topology. The essential feature needed to generate an explosive jet in this model is to store a substantial amount of magnetic free energy within the low-lying closed flux. In configurations with a nearly uniform majority-polarity background field, twisting the internal closed flux by imposing slow, quasi-circular surface flows eventually leads to onset of a kink-like instability. Strong feedback between the ideal triggering mechanism and rapid reconnection through the null-point current sheet releases much of the stored free energy and generates a helical, Alfv\'enic jet \citep{Pariat2009,Pariat2010,Pariat2015,Pariat2016,Wyper2016,Wyper2016b,Karpen2017}. In more recent work, we investigated cases that have a strong majority-polarity concentration adjacent to the minority-polarity intrusion. In that case, a filament channel of strongly sheared, low-lying magnetic flux forms at the polarity inversion line (PIL) between the two concentrations. Reconnection above the PIL forms a flux rope that can support a mini-filament, which rises slowly and eventually erupts as a jet through reconnection between the flux rope and the external field \citep{Wyper2017,Wyper2018}. The underlying mechanism is an exact analogue to the breakout model that explains fast coronal mass ejections \citep{Antiochos1998,Antiochos1999}.

Our preceding jet modelling assumed photospheric rotational motions that were strictly internal to the closed-field region. Consequently, we obtained two types of reconnection-driven outflows: weak intermittent plasma releases, and energetic helical jets accompanied by transient bright points beneath the {domed separatrix}. In this paper, we study the activity driven by larger-scale, linear footpoint motions that transport the minority-polarity intrusion across the solar surface, as occurs with {MMEs} in coronal holes. Specifically, we consider the scenario when this motion subjects the closed field beneath the separatrix to a broad {shear.}
We have investigated three configurations of increasing complexity and activity. 
In the first, the minority-polarity flux was moved quite uniformly, but part of the surrounding majority-polarity background flux was left behind due to a gradient in the imposed surface flow. The separatrix was distorted by this shear flow, inducing the development of currents and low-intensity reconnection at the null point. We show that the reconnection process has a natural periodicity, reminiscent of a bright point with quasi-periodic intensity fluctuations. In the second configuration, the minority-polarity patch was placed closer to the gradient in the imposed photospheric flow,  so that the flux was substantially sheared and the surrounding PIL was strongly distorted. This case exhibited elevated reconnection and explosive jetting. It transitioned between long-duration, low-intensity bright point-like reconnection and {short-duration, high-intensity jet-like} reconnection and back again. Third, we added a concentration of majority-polarity flux to the second configuration, then advected the minority- and new majority-polarity concentrations past one another. This ``fly-by'' configuration produced still more energetic and explosive activity as the MMEs first connected to, then disconnected from, each other as they passed. 

In \S\S \ref{sec:setup} and \ref{sec:diagnostics}, we describe the simulation setup and the diagnostics used to interpret the results. The simulations and our analyses are presented in \S \ref{sec:results}. We demonstrate good quantitative agreement between our results and observations, and discuss the implications of our work for natural links between bright points, CH jets, and plumes, in \S \ref{sec:discussion}. In \S \ref{sec:summary} we briefly list the major conclusions of this work.

\section{Setup}
\label{sec:setup}
\begin{figure}
\centering
\includegraphics[width=0.47\textwidth]{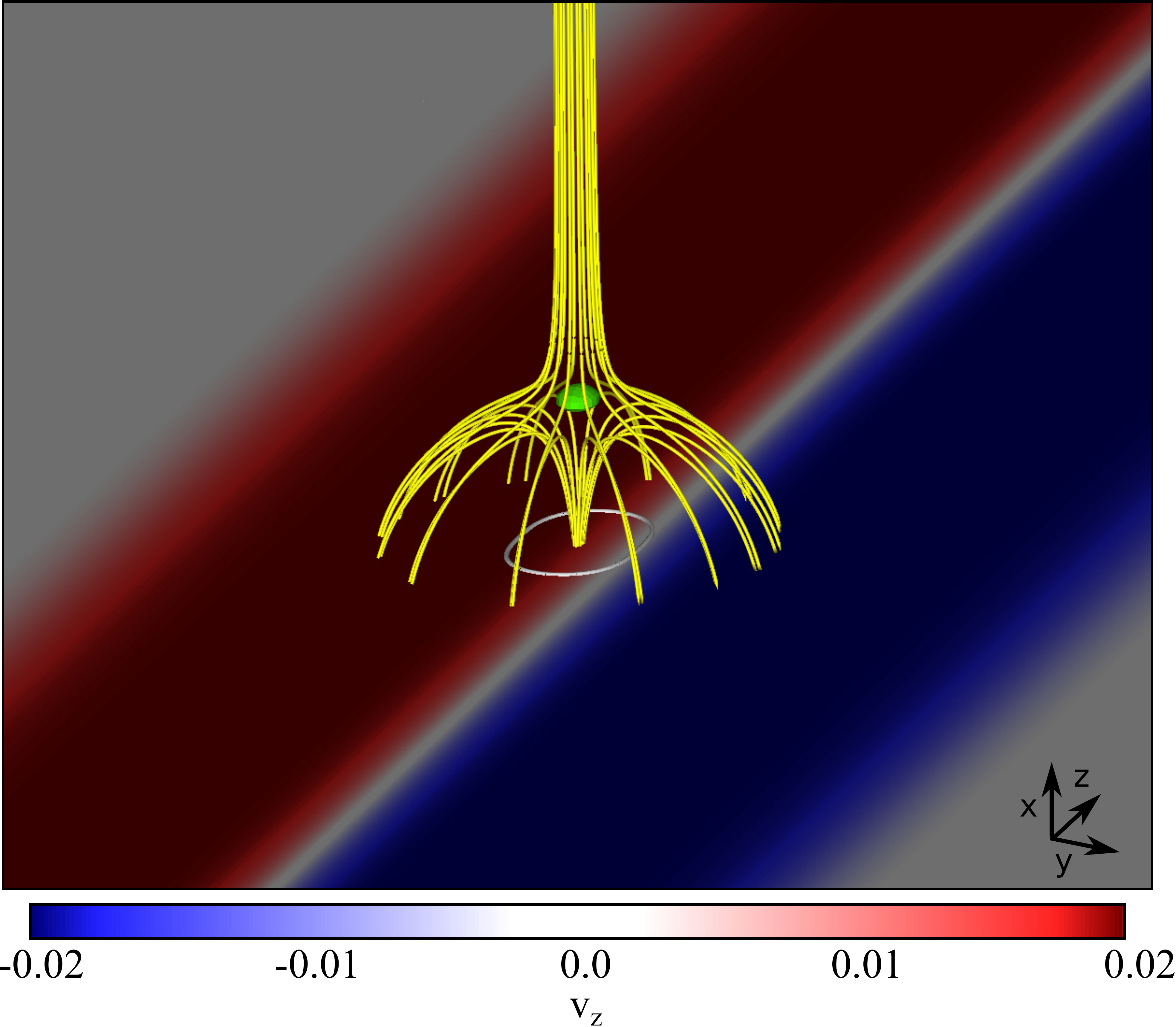}
\caption{The magnetic field and surface flow profile in Configuration 2. Yellow field lines depict the spine-fan topology of the field above the minority polarity. The green sphere is an isosurface of plasma $\beta$ = 10, showing the position of the null point. The PIL is shown as a white contour. Color shading on the surface shows the magnitude and sign of $v_z$.} 
\label{fig:initial}
\end{figure}

We set up a simple system consisting of a single minority-polarity (positive) concentration embedded in a majority-polarity (negative) background. Figure \ref{fig:initial} shows the magnetic field (yellow lines) above the minority polarity in one of the simulations. The field has the typical fan-spine null-point topology associated with an embedded minority polarity \citep[e.g.,][]{Antiochos1996}. Also shown are the surface polarity inversion line (white contour) and a unidirectional imposed flow (color shading) on the surface. For maximum generality, we conducted our experiments in non-dimensionalised units. As discussed below, the results can be scaled to physical units by applying appropriate solar scale factors to the non-dimensionalised results.

We constructed the minority polarity by superposing five vertically aligned, sub-photospheric magnetic dipoles, such that the initial potential field $\bf B$ and vector potential $\bf A$ are given by 
\begin{gather}
\mathbf{B} = \left(b_{0},0,0\right) + \sum_{i=1,5} \boldsymbol{\nabla} \times \mathbf{A}_{i},\\
\mathbf{A}_{i} = \frac{b_{i}{\vert x_{i} \vert}^{3}}{2\left[ x_{i}'^{2} + y_{i}'^{2} + z_{i}'^{2} \right]^{3/2}}\left[ -z_{i}' \hat{\mathbf{y}} + y_{i}' \hat{\mathbf{z}} \right],
\end{gather}
where $x_{i}' = x-x_{i}$, $y_{i}' = y-y_{i}$, and $z_{i}' = z-z_{i}$. We set $b_{0} = -1.0$, $b_{i} = 11.0$, and $x_{i} = -1.0$ in all simulations. The five dipoles are placed on a line along the $z$ direction from $z_{i}=-7.0$ to $z_{i}=-5.0$ at intervals of $\Delta z_{i}=0.5$. $y_{i} = y_{0}$ is a fixed constant that controls the separation between the centre of the minority-polarity concentration ($y=y_0$) and the centre of the surface flow profile ($y=0$).

\begin{figure}
\centering
\includegraphics[width=0.45\textwidth]{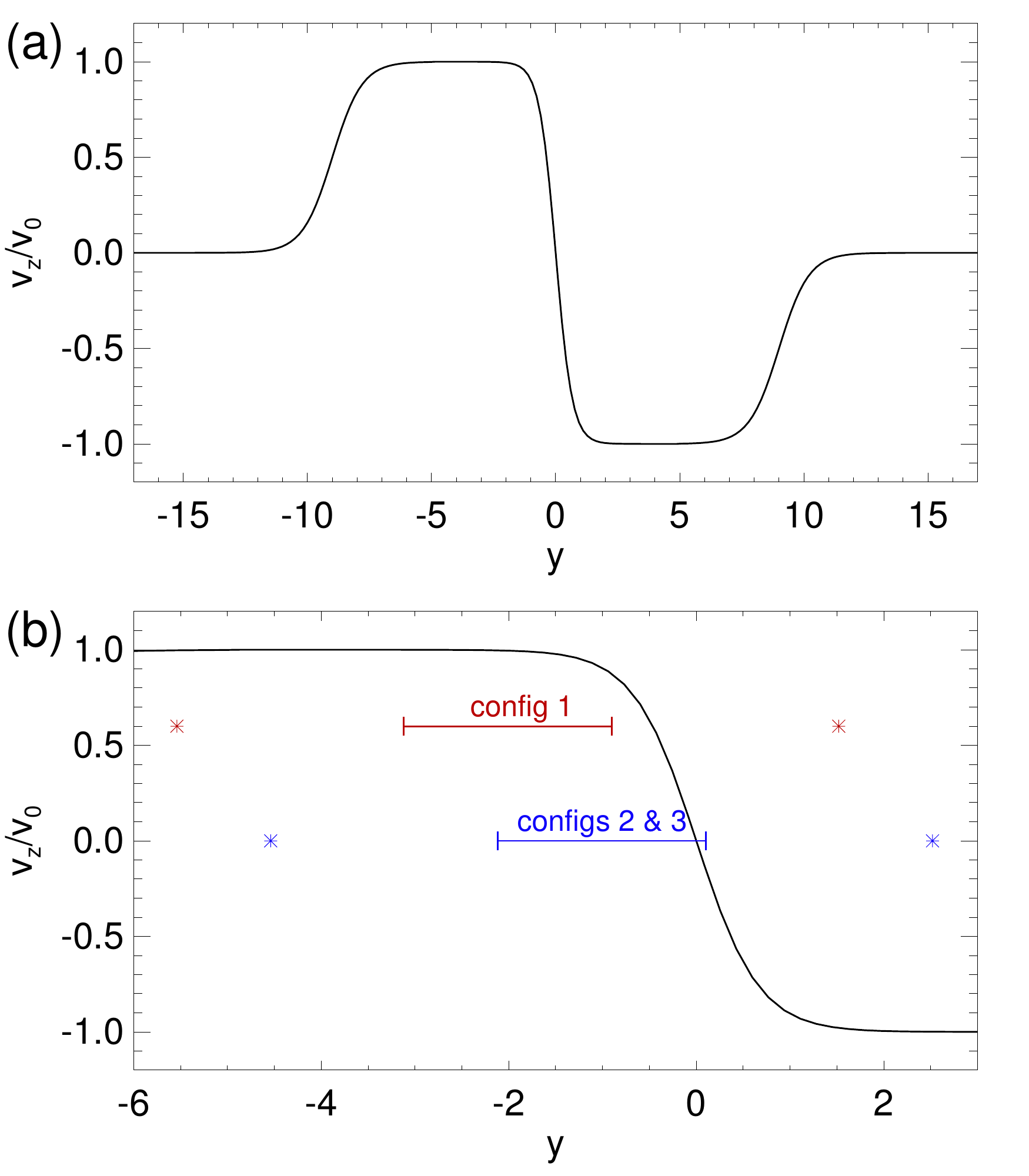}
\caption{(a) Normalized surface flow profile. (b) Zoom-in of (a) near the minority-polarity patch. Blue and red lines show the position and $y$ extent (bounded on the left and right by the PILs) of the minority-polarity patch, and asterisks show the $y$ extent of the separatrix footprint, for Configurations 1, 2, and 3.} 
\label{fig:vel}
\end{figure}

The surface flow is given by
\begin{align}   
v_{z} = \left\{
\begin{array}{cc}
      \frac{v_{0}}{2}\left(\tanh\left(\frac{5(y+9)}{6}\right)+1\right), & -17\leq y< -5; \\
      -v_{0}\tanh\left(\frac{3y}{2}\right), & -5 \leq y < 5; \\
      \frac{v_{0}}{2}\left(\tanh\left(\frac{5(y-9)}{6}\right)-1\right),  & 5 \leq y < 17; \\
      0, & 17 \leq |y|. \\
\end{array} 
\right. 
\end{align}
The profile is shown in Figure \ref{fig:vel}(a). Two bounded regions exist where the surface field is translated bodily along $z$ in opposite directions. These are separated by a shear region of width $w \approx 3$ centred at $y=0$. The driving profile is divergence-free, so that the surface magnetic flux is sheared as it is advected, but it is not cancelled anywhere. 

\begin{figure}
\centering
\includegraphics[width=0.5\textwidth]{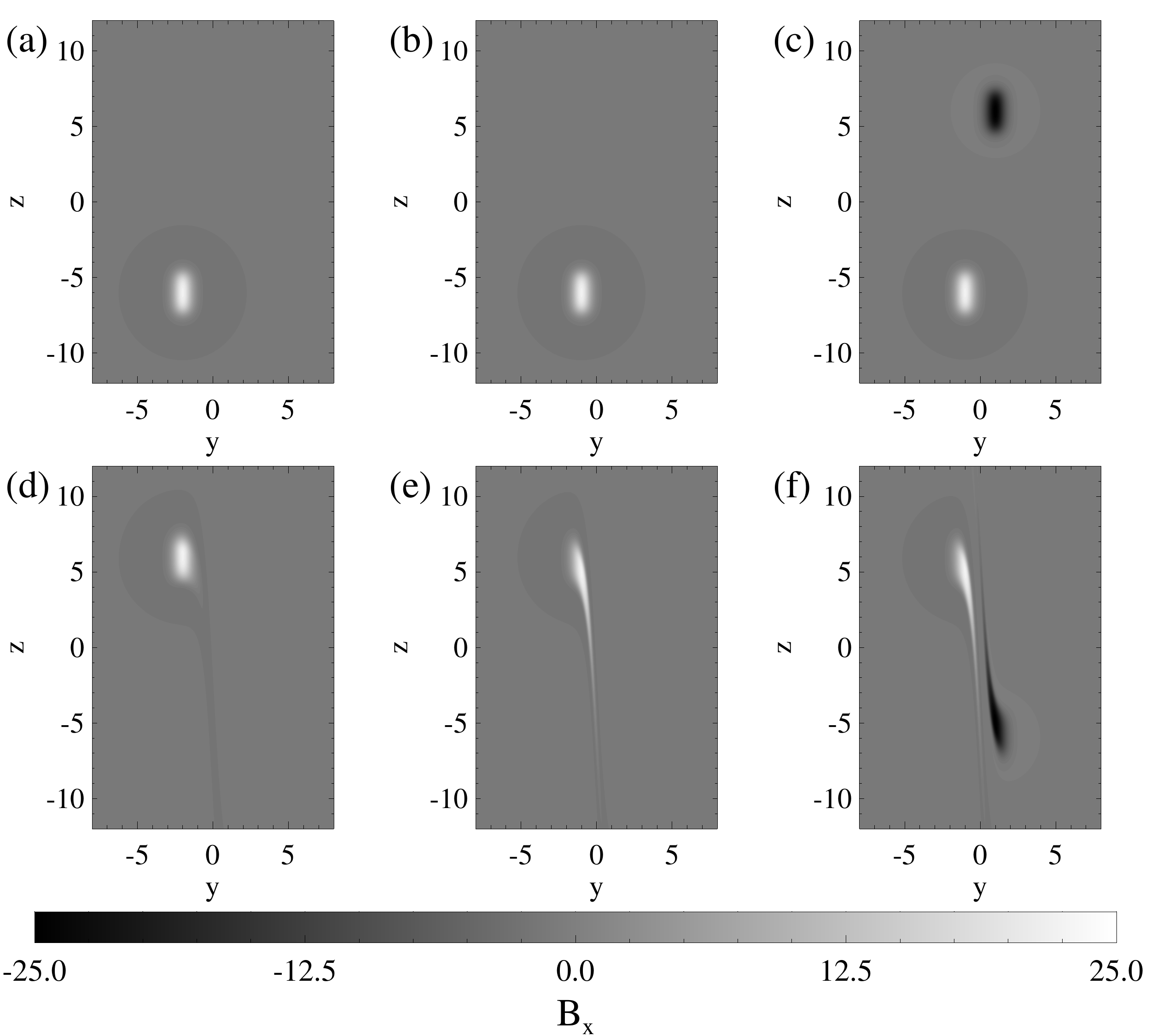}
\caption{Surface $B_{x}$ evolution in the $3$ configurations. Left panels: Configuration 1. Middle panels: Configuration 2. Right panels: Configuration 3. Top row: $t=0$, bottom row: $t \ge 650$ ($t \ge 1250$ for Configuration 1S).} 
\label{fig:bx}
\end{figure}

We studied three configurations that were expected to produce increasing levels of jetting activity. In Configuration 1 we set $y_{0} = -2.0$, which positions the minority polarity almost entirely within a region of uniform translational motion. This is illustrated by the red line in Figure \ref{fig:vel}(b), which indicates the extent of the {PIL} for this case. Nevertheless, the patch is close enough to the shear zone about $y=0$ that some field lines emanating from the patch close down on the far side of the shear zone. The footprint of the separatrix surface, shown by the red asterisks in Figure \ref{fig:vel}(b), spans the shear region, so the separatrix is sheared directly by the flow. Configuration 1, therefore, represents an almost-uniform advection of a minority-polarity MME through weak majority-polarity flux. The initial and final surface flux distributions for this case are shown in Figure \ref{fig:bx}(a,d).

For Configuration 2 we set $y_{0} = -1.0$, so that part of the PIL lies very close to the centre of the shear zone. This is illustrated by Figure \ref{fig:initial} and by the blue line in Figure \ref{fig:vel}(b). The minority-polarity patch itself is sheared strongly by the driving motions in this case, and free energy is injected deep into the closed-field region. Hence, Configuration 2 represents strongly nonuniform advection of the minority-polarity MME. The initial and final surface flux distributions for this case are shown in Figure \ref{fig:bx}(b,e).

Configuration 3 is a modification of Configuration 2 in which we added a strong patch of majority-polarity (negative) flux, equidistant from the centre of the shear zone but in the region of oppositely directed flow. The majority-polarity patch is constructed from five dipoles like the minority polarity, but with $b_{i} = -11.0$, $y_i = +1.0$, and $z_i=+5.0$ to $z_i=+7.0$. It is advected towards and past the minority polarity by the large-scale flow. Configuration 3 thus presents a sheared ``fly-by'' of two MMEs, minority and majority, as shown in Figure \ref{fig:bx}(c,f). We will demonstrate later that their interaction leads to increased jetting activity during the evolution, relative to Configurations 1 and 2. Note that in spite of the extreme deformation of the initial flux distribution, as evident in Figure \ref{fig:bx}(c,f), the normal flux of the minority polarity at the boundary {($\Psi_{pp}\approx 55.4$)} was conserved to within 98\%. 

In all cases, we ramped up the flow over $50$ time units, held it constant for a time, and then ramped it down to zero, again over $50$ time units. The minority-polarity patch was advected from its centred starting position at $z=-6.0$ to $z=+6.0$ (Fig.\ \ref{fig:bx}). We considered two flow speeds, slow (S) and fast (F), for Configuration 1. Configuration 1S had a driving speed $v_{0} = 0.01$ and required a total driving duration of $1250$ time units, whereas Configuration 1F had $v_{0}=0.02$ over a duration of $650$ time units. In  Configurations 2 and 3, we applied only the fast flow of Configuration 1F.

We used the Adaptively Refined Magnetohydrodynamics (MHD) Solver \citep[ARMS;][]{DeVore2008} to solve the ideal MHD equations in the form
\begin{gather}
\frac{\partial \rho}{\partial t} + \boldsymbol{\nabla}\cdot(\rho \mathbf{v}) = 0,\\
\frac{\partial (\rho \mathbf{v})}{\partial t}+\boldsymbol{\nabla}\cdot(\rho \mathbf{v}\mathbf{v}) + \boldsymbol{\nabla} P -\frac{1}{\mu_{0}}(\boldsymbol{\nabla}\times\mathbf{B})\times\mathbf{B} = 0,\\
\frac{\partial U}{\partial t}+\boldsymbol{\nabla}\cdot (U\mathbf{v})+P\boldsymbol{\nabla}\cdot\mathbf{v} = 0,\\
\frac{\partial \mathbf{B}}{\partial t}-\boldsymbol{\nabla}\times (\mathbf{v}\times\mathbf{B})=0,
\end{gather}
where $t$ is the time, $\rho$ is the mass density, $P = \rho R T$ is the thermal pressure, $U = P/(\gamma-1)$ is the internal energy density, $\mu_{0}=4\pi$ is the magnetic permeability, and $\mathbf{B}$ and $\mathbf{v}$ are the 3D magnetic and velocity fields. An ideal gas is assumed, with ratio of specific heats $\gamma = 5/3$. Similar to our previous jet experiments \citep[e.g.,][]{Wyper2016,Wyper2018}, we imposed an initially uniform plasma density, temperature, and pressure of $1.0$, $1.0$, and $0.01$ respectively; hence, the non-dimensional gas constant is $R=0.01$. The corresponding plasma $\beta \approx 2\times 10^{-1}$ in the background field, dropping to $\beta \approx 5\times 10^{-4}$ at the surface within the minority polarity. The initially uniform sound speed $v_{s} \approx 0.13$, whilst the Alfv\'{e}n speed varies from $v_{a} \approx 0.3$ in the background field to $v_{a} \approx 6.5$ in the minority polarity. The largest driving speed used in our simulations is $0.02$, approximately $15\%$ of the sound speed and $6.7\%$ of the background Alfv\'{e}n speed. Thus, the region around our moving magnetic elements evolved quasistatically in response to the imposed surface flow. Reconnection eventually occurred in all of our simulations due to inherent numerical diffusion during the time advancement of the MHD equations, in particular the induction equation for the magnetic field. {This effective numerical magnetic diffusion is as small as possible for the given simulation grid whilst maintaining numerical stability and monotonicity \citep{DeVore1991}.}

When the driving began in the two counter-propagating flow regions, two large-scale Alfv\'{e}n wave fronts propagated upwards into the simulation volume. The box size in each simulation is $[0,192]\times[-32,32]\times[-32,32]$, tall enough that the disturbance from the driving profile reached the top boundary at $t \approx 680$. The simulations with $v_{0}=0.02$ were halted at about this time. Consistent with the boundary driving, periodic boundary conditions were applied to the $z$ boundaries, whilst the side $y$ boundaries were closed and line-tied. Open, zero-gradient boundary conditions were applied at the top boundary, except that the tangential velocity components were set to zero beyond the boundary to partially damp the slippage of field lines. Some of the Alfv\'{e}n-wave disturbance was reflected from this boundary in the case with $v_{0}=0.01$, where the driving was applied over a longer time. However, the simulation was halted before the reflections reached the jetting region at $t\approx 1360$. The bottom boundary was closed and line-tied everywhere. Except where the driving profile above was prescribed, the tangential velocity was zero.

The simulation grid was adapted dynamically and managed by the PARAMESH toolkit \citep{MacNeice2000}. It refined/de-refined according to local measures of the gradient and strength of the magnetic field \citep[see the Appendix in][]{Karpen2012}. In terms of the adaptive parameters introduced in \citet{Karpen2012}, we used $c_{1} = 0.01$, $c_{2} = 0.04$, $B_{1} = 1\times 10^{-4}$, and $B_{2} = 20$. These values were found to resolve the reconnection region and jet outflows well, and to track them as they propagated upwards into the simulation volume. We used five levels of grid refinement in these simulations, corresponding to a minimum grid spacing of $6.25\times 10^{-2}$. Additionally, a sub-volume large enough to encompass the separatrix surface as it was advected across the domain was fixed at the maximum refinement level throughout the simulations, to maximally resolve the dynamics around the MMEs.

For generality, the equations were solved in dimensionless form. Hence, the time units of the simulations can be understood relative to characteristic time scales of the system. The separatrix dome is $\approx 7$ length units wide, whilst the Alfv\'{e}n speed at the centre of the minority polarity is $\approx 6.5$ velocity units. Thus, a time unit of $1$ corresponds roughly to the travel time of an Alfv\'{e}n wave across the width of the separatrix dome. Our results can be scaled to solar observations by fixing a typical dome length scale ($L_{s}$), magnetic-field strength ($B_{s}$), and plasma density ($\rho_{s}$). Other useful quantities such as 
\begin{align}
V_{s} = \frac{B_{s}}{\sqrt{\rho_{s}}}, \quad t_{s} = \frac{L_{s}}{V_{s}}, \quad E_{s} = B_{s}^2 L_{s}^3
\end{align}
that fix the velocity, time, and energy scales then can be deduced. For example, choosing $L_{s}$ = $2.5 \times 10^8$\,cm, $B_{s}$ = 2.5\,G, and $\rho_{s}$ = $4 \times 10^{-16}$\,g cm$^{-3}$ gives $V_{s}$ = $1250$\,km\,s$^{-1}$, $t_{s}$ = 2\,s, and $E_{s}$ = $9.8 \times 10^{25}$ erg, yielding impulsive speeds, time scales, and liberated energies typical of coronal bright points and jets ({see \S \ref{sec:discussion}}). For simplicity and generality, however, the values quoted below are given in non-dimensionalised units. 

\begin{figure}
\centering
\includegraphics[width=0.48\textwidth]{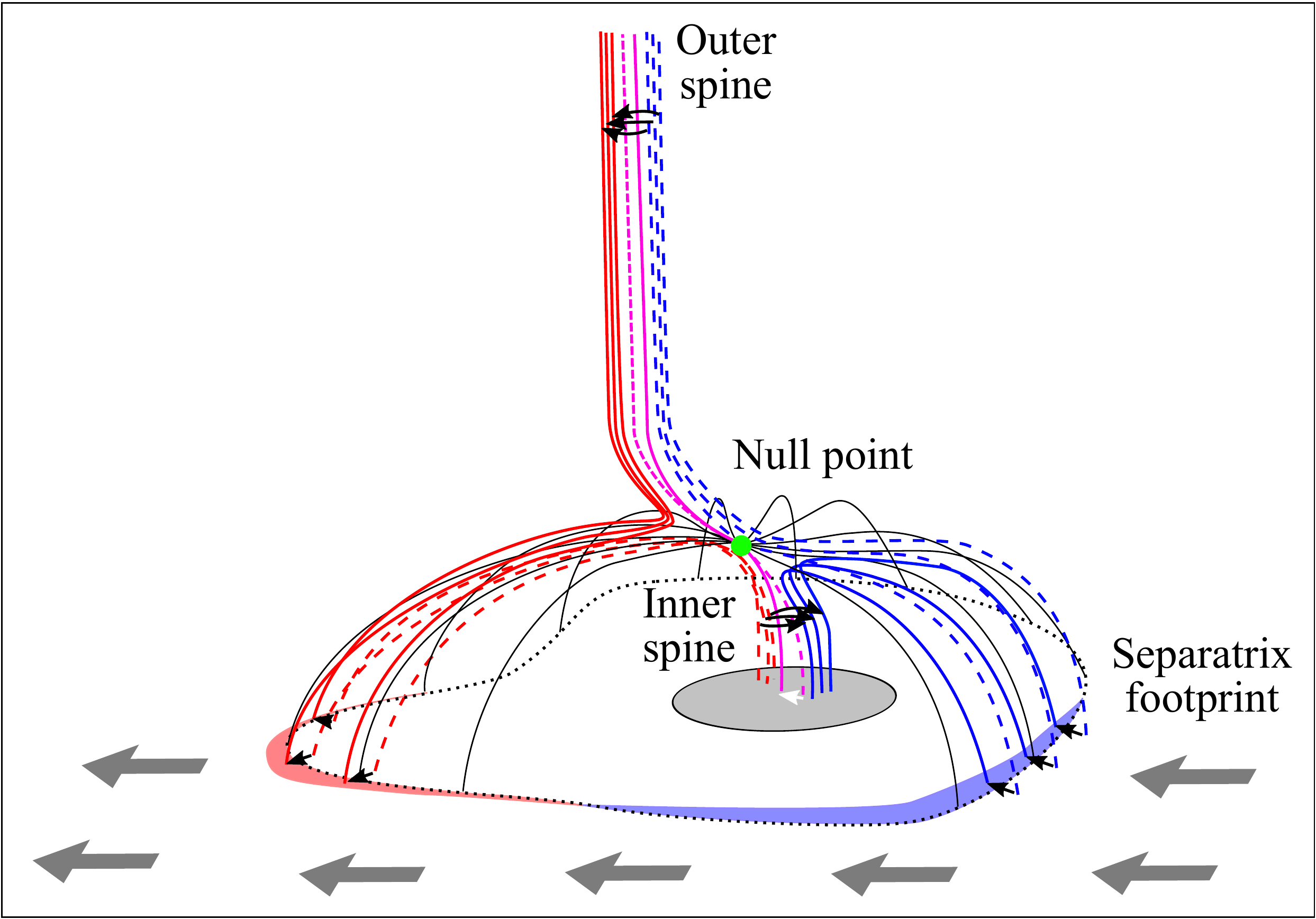}
\caption{Schematic of the initial flux transfer process; see text for details. The fat grey arrows symbolize the surface flow in the reference frame of the moving minority-polarity element. The small black arrows indicate the direction of flux transfer: from dashed to solid across the separatrix and across/around the spine lines. The white arrow indicates the movement of the inner spine: from dashed to solid pink.}
\label{fig:schematic}
\end{figure}

\begin{figure*}
\centering
\includegraphics[width=1.0\textwidth]{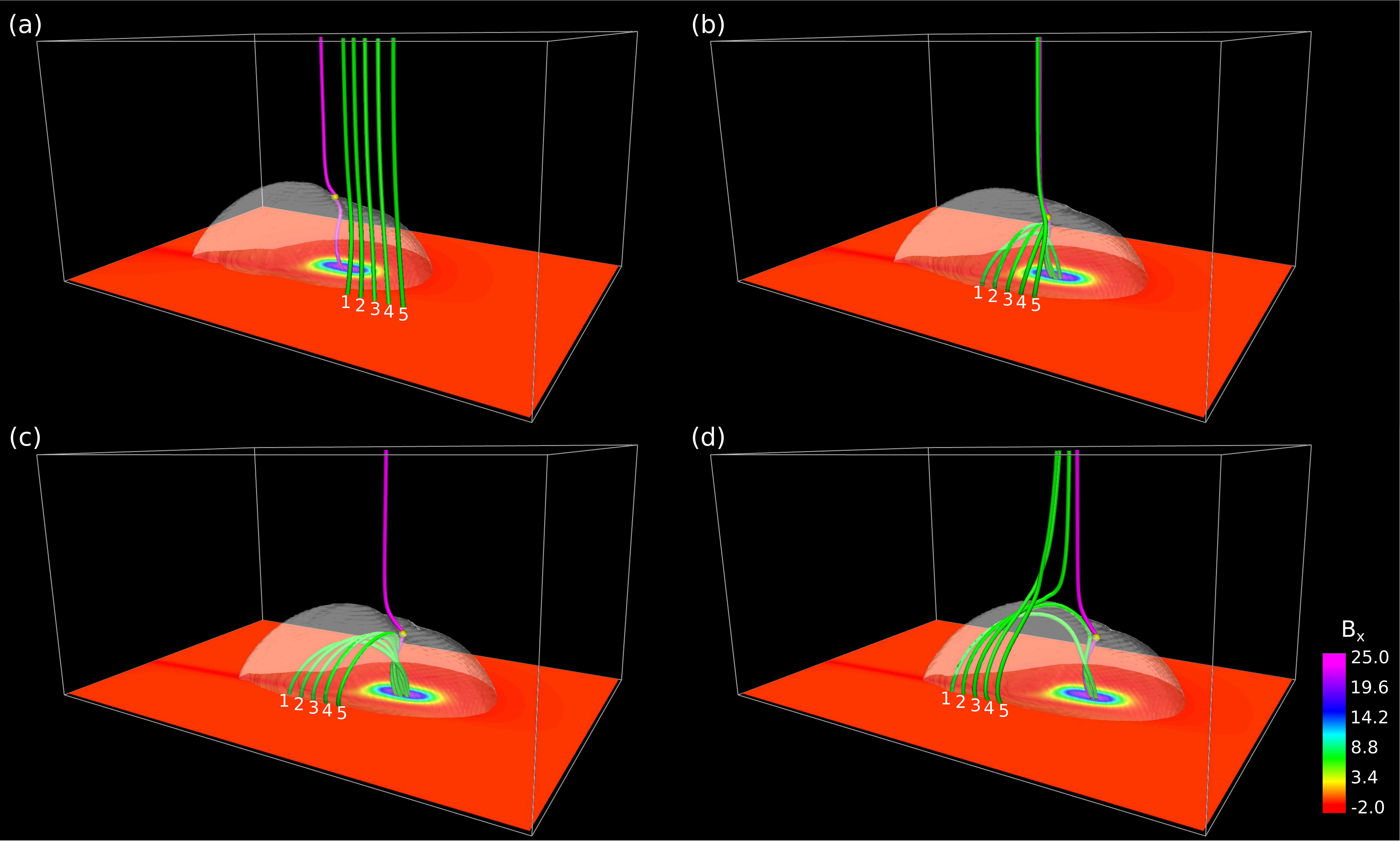}
\caption{Field lines reconnecting during one cycle in Configuration 1S. Semi-transparent isosurfaces show the separatrix, magenta field lines mark the spine, and the yellow sphere is the null point. Green field lines are traced from footpoints on the surface that move along with the surface motions. Times are (a) $t = 590$, (b) $t = 770$, (c) $t = 870$, and (d) $t = 940$. An animation is available online.}
\label{fig:simcomp}
\end{figure*}

\begin{figure*}
\centering
\includegraphics[width=1.0\textwidth]{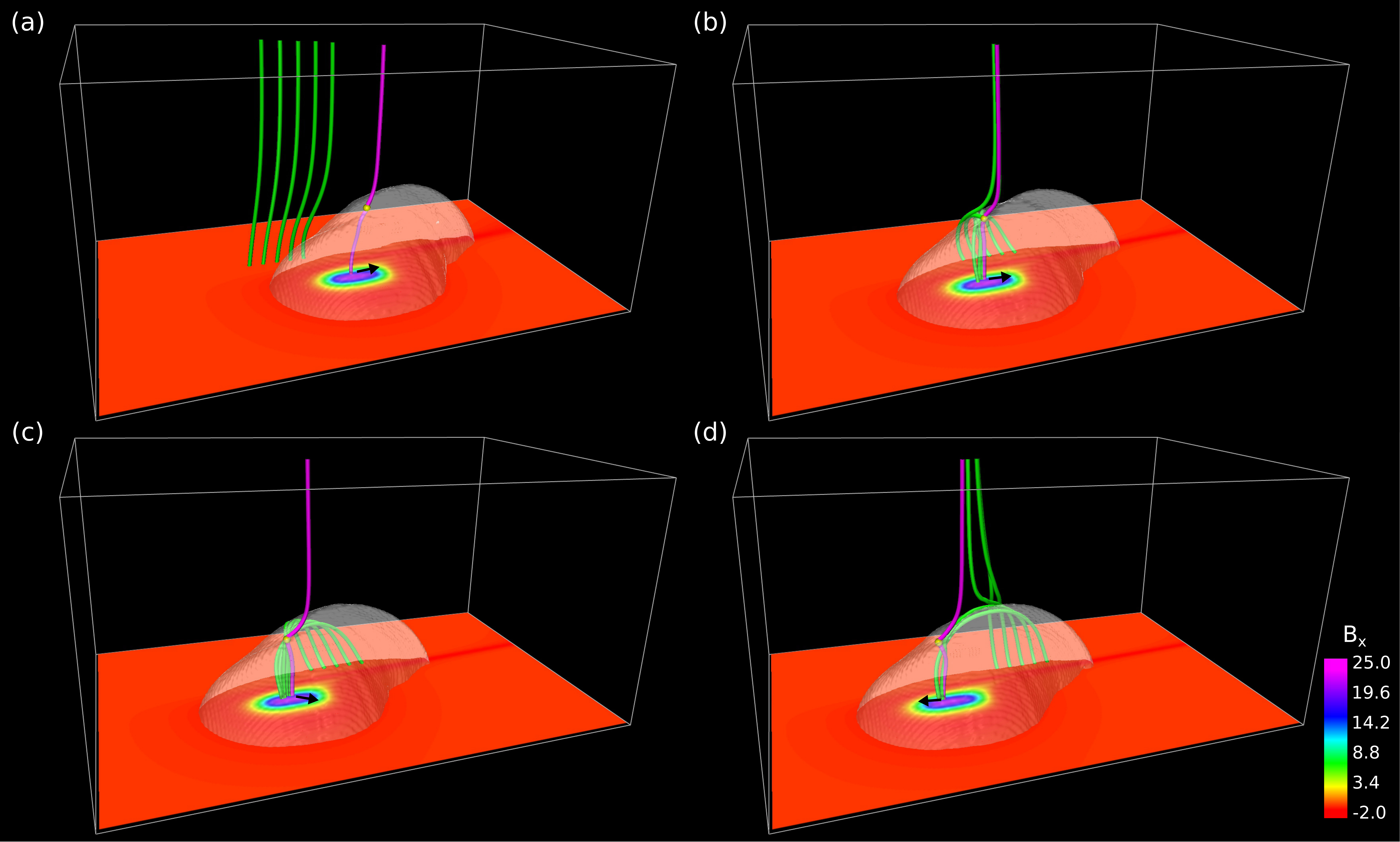}
\caption{Movement of the footpoint of the inner spine during one cycle in Configuration 1S. Quantities and times shown are the same as in Figure \ref{fig:simcomp}, but viewed from the opposite side of the domain. The black arrows indicate the instantaneous direction of motion of the inner spine. An animation is available online.}
\label{fig:simcomp2}
\end{figure*}

\section{Diagnostics}
\label{sec:diagnostics}
For investigating the reconnection process, we identified and tracked the positions of the null point (or cluster of null points) associated with the minority polarity using the tri-linear method described in \citet{Haynes2007}; for details of the implementation in ARMS see the Appendix in \citet{Wyper2016b}. {As in that paper, we also identified the open-closed separatrix associated with the null (or nulls) by sampling the connectivity of points within the domain, and assigning each point a functional value of $f(x,y,z) = 0$ in open field and $f(x,y,z) = 1$ in closed field. The separatrix was then visualised as an isosurface of $f(x,y,z) = 0.5$.} Another informative reconnection diagnostic was the normalised interchange reconnection rate,
\begin{align}
\frac{d\Psi}{dt}(t) \approx \frac{|\Psi_{opened}| + |\Psi_{closed}|}{2 \Psi_{pp}\Delta t}, 
\end{align}
where $\Delta t$ is the time between simulation snapshots, $\Psi_{pp} \approx 55.4$ is the flux of the minority polarity, and $\Psi_{opened}$ and $\Psi_{closed}$ are the amounts of newly opened and closed flux, respectively, since the previous snapshot. Formally, $\Psi_{opened}$ should equal $\Psi_{closed}$, but numerically small variations exist so we take their average. To calculate $\Psi_{opened}$ and $\Psi_{closed}$, the ideal motions imposed at the boundaries must be filtered out. We do this by tracing field lines from a uniform grid on the photosphere at a given time to produce an array of field-line end points, $(X_{t}(y,z),Y_{t}(y,z),Z_{t}(y,z))$. The same starting points are then advected by the boundary flow backwards in time to the previous snapshot (at $t - \Delta t$), whereupon field lines are traced from these positions at this time, giving a second set of field-line end points, $(X_{t-\Delta t}(y,z),Y_{t-\Delta t}(y,z),Z_{t-\Delta t}(y,z))$. This second set is then advected forwards in time to the time of the original snapshot ($t$), whereupon they share the same starting positions as the original mapping, but preserve the connectivity of the previous time. Comparing the two mappings at time $t$, we identify newly opened and newly closed field lines over the time interval $\Delta t$ \citep[for an in-depth discussion, see][]{Titov2009}. Summing over all reconnected field lines, weighted by the magnetic flux at the field-line starting points, then gives $\Psi_{opened}$ and $\Psi_{closed}$ \citep[for further details, see][]{Wyper2016}.

To investigate the energetics of our simulations, we calculated 
\begin{align}
K = \iiint{\frac{1}{2}\rho v^2 \,dx\,dy\,dz}, \\
M = \iiint{\frac{1}{8\pi} B^2 \,dx\,dy\,dz},
\end{align}
the volumetric integrals of kinetic ($K$) and magnetic ($M$) energy, respectively. However, we found that the large-scale Alfv\'{e}n wave disturbances launched by the boundary driving dominated the plots of these quantities, injecting large and sustained increases of kinetic and magnetic energy into the volume. In addition, the boundary shearing motions altered the surface flux distribution, changing the minimum-energy (potential) state of the magnetic field.

To minimize the Alfv\'{e}n-wave effect on the diagnostics, we ran reference simulations identical to each calculation but with the uniform background field only (i.e., omitting the sub-photospheric dipoles), and calculated the energy associated with the large-scale Alfv\'{e}n wave fronts. To account for the minimum-energy effect, we performed a potential-field extrapolation at each time using the evolved surface magnetic flux, thereby computing the changing baseline value of the magnetic energy. By subtracting these contributions from $K$ and $M$, we estimated the kinetic energy ($\Delta K$) generated by the reconnection outflows/jets and the free magnetic energy ($\Delta M_{free}$) related to the stress in the closed-field region, 
\begin{align}
\Delta K &= K - K_{ref}, \\
\Delta M_{free} &= \Delta M - \Delta M_{ref} - \Delta M_{pot}, 
\end{align}
where 
\begin{align}
\Delta M &= M - M(t=0), \\
\Delta M_{ref} &= M_{ref} - M_{ref}(t=0), \\
\Delta M_{pot} &= M_{pot} - M_{pot}(t=0).
\end{align}
$M_{ref}$ and $K_{ref}$ are the kinetic and magnetic energy integrals calculated for the reference simulations, and $M_{pot}$ is the magnetic energy of the potential field associated with the evolving surface flux at the simulation bottom boundary, all evaluated at time $t$. Note that $\Delta M_{free}$ and $\Delta K$ are directly comparable, as $K_{ref}(t=0) = K(t=0) = 0$ and the minimum state for kinetic energy is $K_{pot} = 0$ at all times $t$. 

\begin{figure*}
\centering
\includegraphics[width=\textwidth]{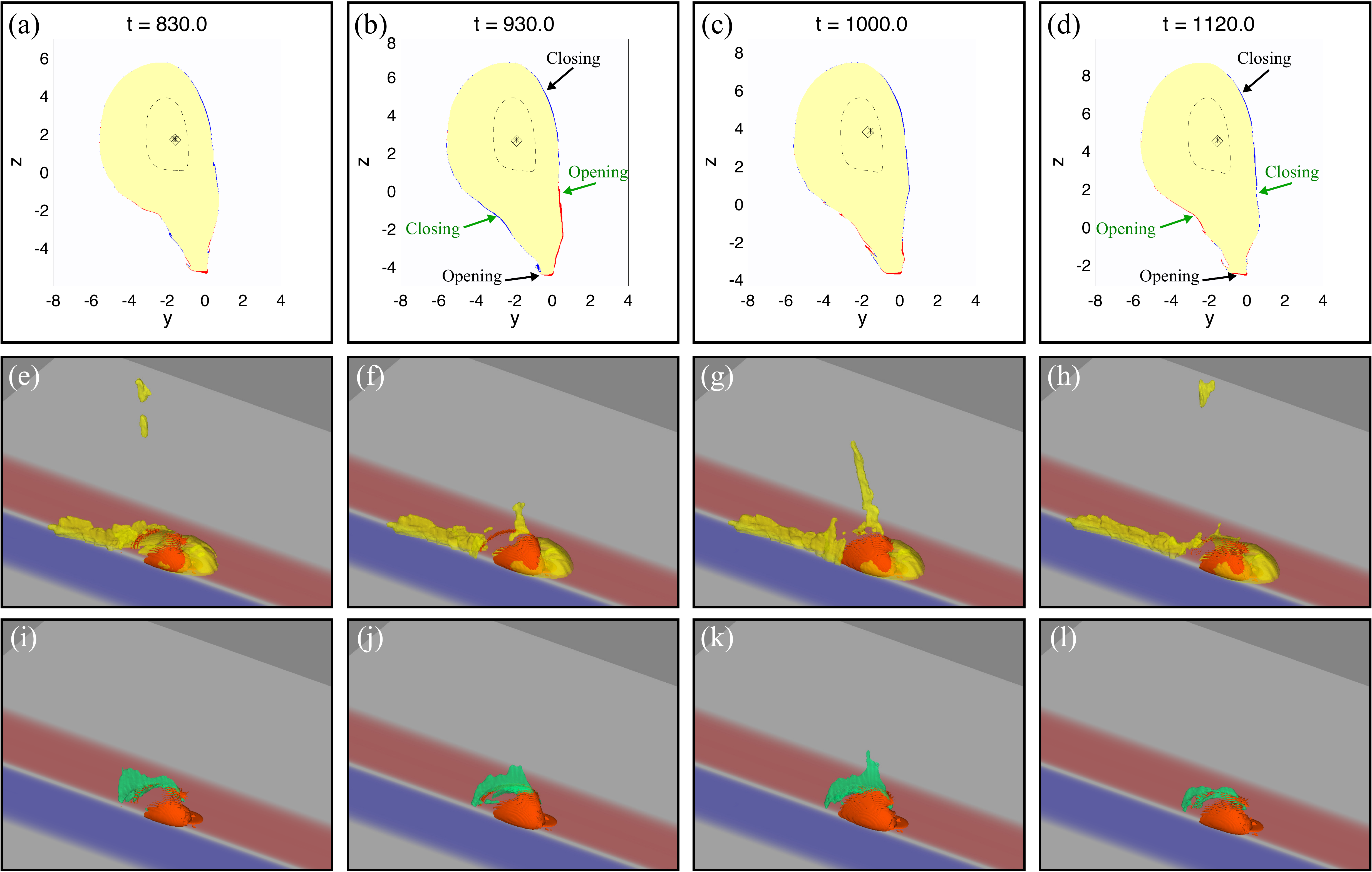}
\caption{Configuration 1S: outflows and reconnected flux during one cycle. Top panels illustrate the connectivity of magnetic flux on the photosphere at the times indicated. White denotes open-field regions, yellow closed-field regions. Blue and red show flux that has recently closed and opened, respectively. The PIL is shown as a dashed line. Asterisks show the projections of the null point positions, whilst the diamond marks the position of the null point centroid (smoothed in time). Middle and bottom panels show the enhanced mass density (yellow isosurface,  $\rho = 1.1$) and plasma velocity (green isosurface, $v = 0.05$), respectively, in the reconnection outflows. Orange isosurfaces show enhanced current density ($J = 0.3$). Red/blue shading shows the surface plasma velocity $v_z$, colour scale as in Figure \ref{fig:initial}(b). An animation of this figure is available online.}
\label{fig:topology1}
\end{figure*}

\begin{figure*}
\centering
\includegraphics[width=0.9\textwidth]{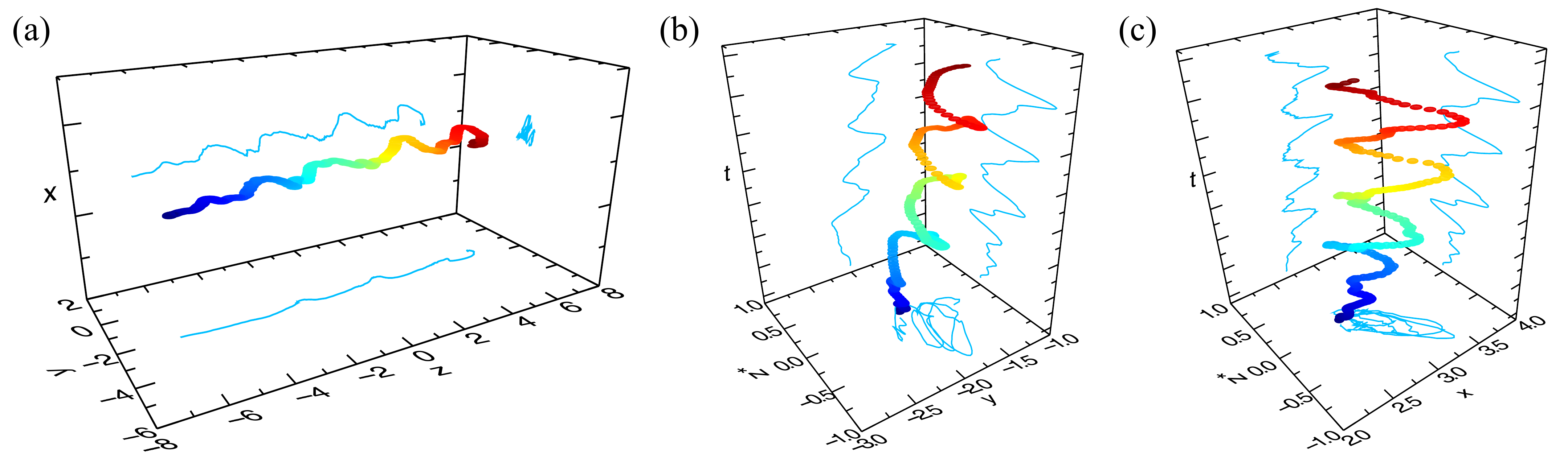}
\caption{Smoothed evolution of the null-point centroid in Configuration 1S. (a) 3D visualisation and 2D projections of the evolving null position. Time increases from dark blue ($t=0$) to dark red ($t=1350$). Note the periodic change of position in $y$ and $x$. (b) Rotation of the centroid in $y$ and $z$ about a central point moving with the flow. The vertical axis is time and $z^*$ is the $z$ variation in the moving coordinate system. (c) same plot but for $x$ vs $z^*$.}
\label{fig:centroid}
\end{figure*}

\begin{figure}[t]
\centering
\includegraphics[width=0.5\textwidth]{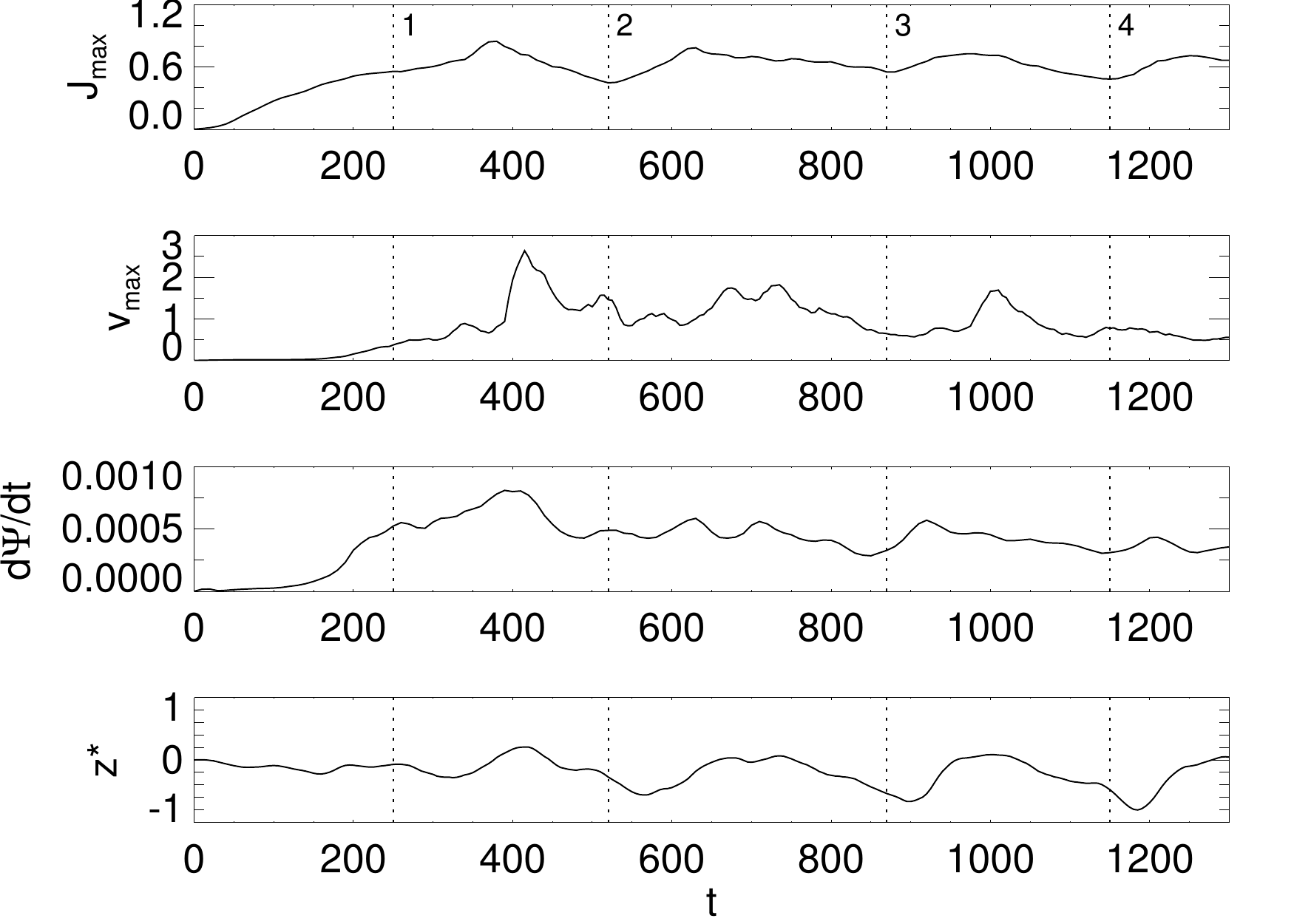}
\caption{Configuration 1S. $J_{max}$, $v_{max}$, $d\Psi/dt$, and $z^*$, the location of the null centroid along the $z$ direction relative to the moving coordinate system. Vertical dotted lines roughly delineate four cycles of energy buildup and release. All quantities have been smoothed.}
\label{fig:slowplots}
\end{figure}

\section{Results}
\label{sec:results}
\subsection{Configuration 1 -- Uniform Advection}
\label{sec:uniform}
In Configuration 1, the flow field was nearly uniform across the minority-polarity concentration, so that closed field near the PIL was only weakly sheared. Part of the footprint of the separatrix surface, on the other hand, was significantly stressed by the driving motions. Consequently, we expect that a current layer will form readily around the magnetic null point due to the displacement of the inner (closed) and outer (open) flux systems, facilitating reconnection there \citep[e.g.,][]{Antiochos1996, Karpen1996, Pontin2007, Edmondson2010}. The fields involved are relatively weak and large scale and they accumulate little free energy, so the reconnection is not expected to be explosive. Instead, because the injected stress will continually propagate towards and be processed through the null point, we anticipate that the reconnection will involve only weak energy release and eventually establish a quasi-steady or quasi-cyclical evolution. 

Figure \ref{fig:schematic} shows a schematic of this process {in a frame of reference moving with the minority polarity}. The minority-polarity region is shown in grey, with the PIL outlined in black. The green dot denotes the null point, whilst solid black and solid/dashed magenta field lines show part of the fan separatrix and spine lines before and after reconnection, respectively. {The thick grey arrows show the surface flows relative to the minority-polarity region.} The action of the {surface flows} drags the separatrix out behind the minority polarity, distorting the footprint of the separatrix on the surface (black dotted line). This stresses the null-point region, collapsing the spine and fan towards each other and forming a current layer around the null, even though the open outer spine is free to move. Reconnection within the current layer induces flux transfer across the separatrix, closing down the arriving flux (blue field lines, moving from dashed to solid) at the front of the separatrix and opening up the departing closed flux (red field lines moving from dashed to solid) at the trailing end of the separatrix. The shaded red and blue regions on the surface indicate the regions over which this flux transfer occurs. 

Flux transfer in this manner is precisely what we see at the start of our simulation of Configuration 1, and is similar to that found in previous investigations of null-point reconnection \citep[e.g.,][]{Edmondson2010,Masson2012,Pontin2013}. Unlike this previous work, however, in our simulation the driving produced a {sustained} deformation of the closed flux region and, consequently, a {continual} storage of free magnetic energy. In order for interchange reconnection {to release} this internal free energy, the null and separatrix must evolve so as to penetrate into the {affected} closed flux region, which generally requires more {intense reconnection dynamics}, similar to what is seen in the standard models for coronal jets. As with homologous jets, {this} continued driving of Configuration 1 resulted in {\it{cycles}} of energy buildup due to the ideal stressing and rapid release by reconnection.

One such reconnection cycle is shown from two viewpoints in Figures \ref{fig:simcomp} and \ref{fig:simcomp2}. The green field lines are traced from footpoints on the surface that move with the driving flows and are sequentially reconnected (from 1 to 5) into the minority polarity at the front of the separatrix (semi-transparent isosurface) as the cycle begins. Note that the associated newly opened field lines at the back of the separatrix are omitted for clarity. As the green field lines are reconnected at the null, the field lines are connected sequentially on to the parasitic polarity patch. Each new connection is made further away from the leading edge of the separatrix than the last. At the same time, the inner-spine footpoint follows these connections and is shifted backwards across the minority polarity (Fig. \ref{fig:simcomp2}(a)-(b)). As the driving continues, the surface motions drag the other footpoint of each field line past the minority polarity, folding the first field lines to be reconnected underneath the later ones and introducing a half turn of twist within this region (Fig. \ref{fig:simcomp}(b)-(c)). The slight increase in magnetic pressure associated with the newly closed flux compresses the underside of the null as it passes the minority polarity, changing the angle of the current layer around the null.  Consequently flux begins to open all along the side of the separatrix closer to the center of the domain, shifting the inner-spine footpoint sideways (Fig. \ref{fig:simcomp2}(c)). A burst of reconnection occurs when this magnetic flux reaches the trailing edge of the separatrix surface, whereupon the inner-spine footpoint moves back to where it started (Fig. \ref{fig:simcomp2}(d)) as the closed field lines reconnect again in reverse order (Fig. \ref{fig:simcomp}(d)). In the animation accompanying Figure \ref{fig:simcomp} a whip-like relaxation of the last field lines to open can be seen, accelerating plasma in the reconnection outflow (discussed further below).

Throughout the cycle, magnetic flux continually closed down at the front of the separatrix and opened at the trailing end; no reversal of the main reconnection direction occurred as in the so-called ``oscillatory reconnection'' scenario \citep[e.g.,][]{Craig1991,Thurgood2017}. Rather, reversals of flux transfer took place along the sides of the separatrix. One such reversal is shown in Figure \ref{fig:topology1}(a)-(d), which displays the connectivity of the surface flux at various times during the same cycle. The flux transfer on the lower right side varies from blue to red (closing to opening) and back again, whilst the front and back of the separatrix surface show continual closing (blue) and opening (red) of flux respectively. Additionally, the lower left side of the separatrix also undergoes the opposite flux reversal to the right side (red to blue and back again) to preserve the net flux that is opened/closed. 

Along with the footpoint of the inner spine, the null point also precessed in a circle as the closed-field evolution changed the stresses around the null. On top of this periodic motion, the null reconnection region was also slightly {fragmented}, which added some unsteadiness to the outflows and intermittently replaced the original null point with a small cluster of null points. To focus on the cyclic behaviour, we show the position of the null point centroid (the average position of all nulls at a given time) after smoothing in Figure \ref{fig:centroid}(a). The periodicity of its position as the minority polarity moved in the positive $z$ direction is somewhat evident. However, when we shift to a frame of reference where the minority polarity is stationary ($z \to z^*$), the cyclic change in the null position becomes clear (Fig. \ref{fig:centroid}(b)-(c); see also the animation of Fig.\ \ref{fig:topology1}).

The duration and frequency of the reconnection cycles {are displayed} in Figure \ref{fig:slowplots}, which shows the peak current in the current layer, $J_{max}$  (evaluated for field strengths where $B \le 2.0$ to discount high volumetric currents), the peak plasma velocity in the volume, $V_{max}$ (equivalent to the Alfv\'{e}nic reconnection outflow velocity close to the current layer), the interchange reconnection rate, $d\Psi/dt$, and the position of the null in the moving coordinate system, $z^*$ . Although the peak plasma velocity was significantly higher than the eventual outflowing plasma speed along the open field lines, which is closer to the local sound speed, it remains a useful indicator of the presence of reconnection-driven flows.
 
Following a ramp-up phase ($t \le 250$), the system underwent three clear reconnection cycles and the start of a fourth before the driving began to ramp down at $t = 1200$. In each cycle, roughly the same changes in each quantity occurred, with peaks in current density and reconnection rate followed shortly thereafter by a peak in plasma velocity caused by a burst of outflow and a forward shift of the null position. However, significant variations among cycles are evident, as the burst of reconnection does not produce a simple reproducible evolution in which each cycle brings the system back to its initial pre-stressed state.  The interchange reconnection rate and the peak current around the null evolved differently, as flux is transferred across the current sheets spanning the separatrix, not just at the null itself \citep[e.g.,][]{Pontin2005,Wyper2012}. Variations in the interchange reconnection rate slightly precede the corresponding variations in the peak plasma velocity, because the plasma is not instantaneously accelerated by the Lorentz force of newly reconnected field lines. The final outflow burst, shown in Figure \ref{fig:topology1}(g) and (k), formed a narrow plasma spire that was denser and faster than the preceding quasi-steady reconnection flows. A similar spire formed during each cycle, seen clearly in the online animation of the figure.

\begin{figure}[t]
\centering
\includegraphics[width=0.5\textwidth]{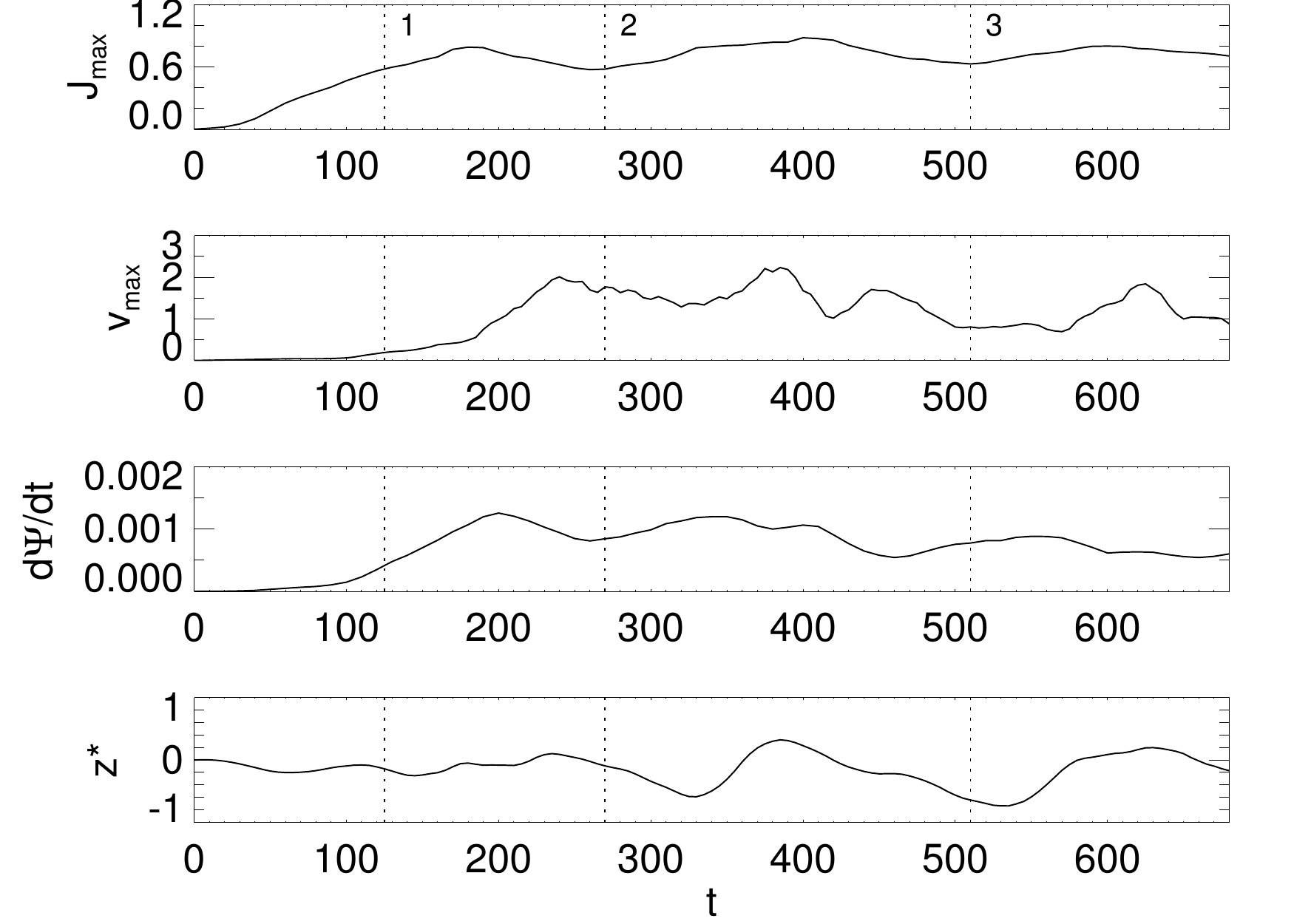}
\caption{Configuration 1F. $J_{max}$, $v_{max}$, $d\Psi/dt$, and $z^*$, the motion of the null centroid in the $z$ direction relative to the moving coordinate system. Vertical dotted lines roughly delineate three cycles of energy buildup and release. All quantities have been smoothed.}
\label{fig:bodyplots}
\end{figure}

\begin{figure}
\centering
\includegraphics[width=0.47\textwidth]{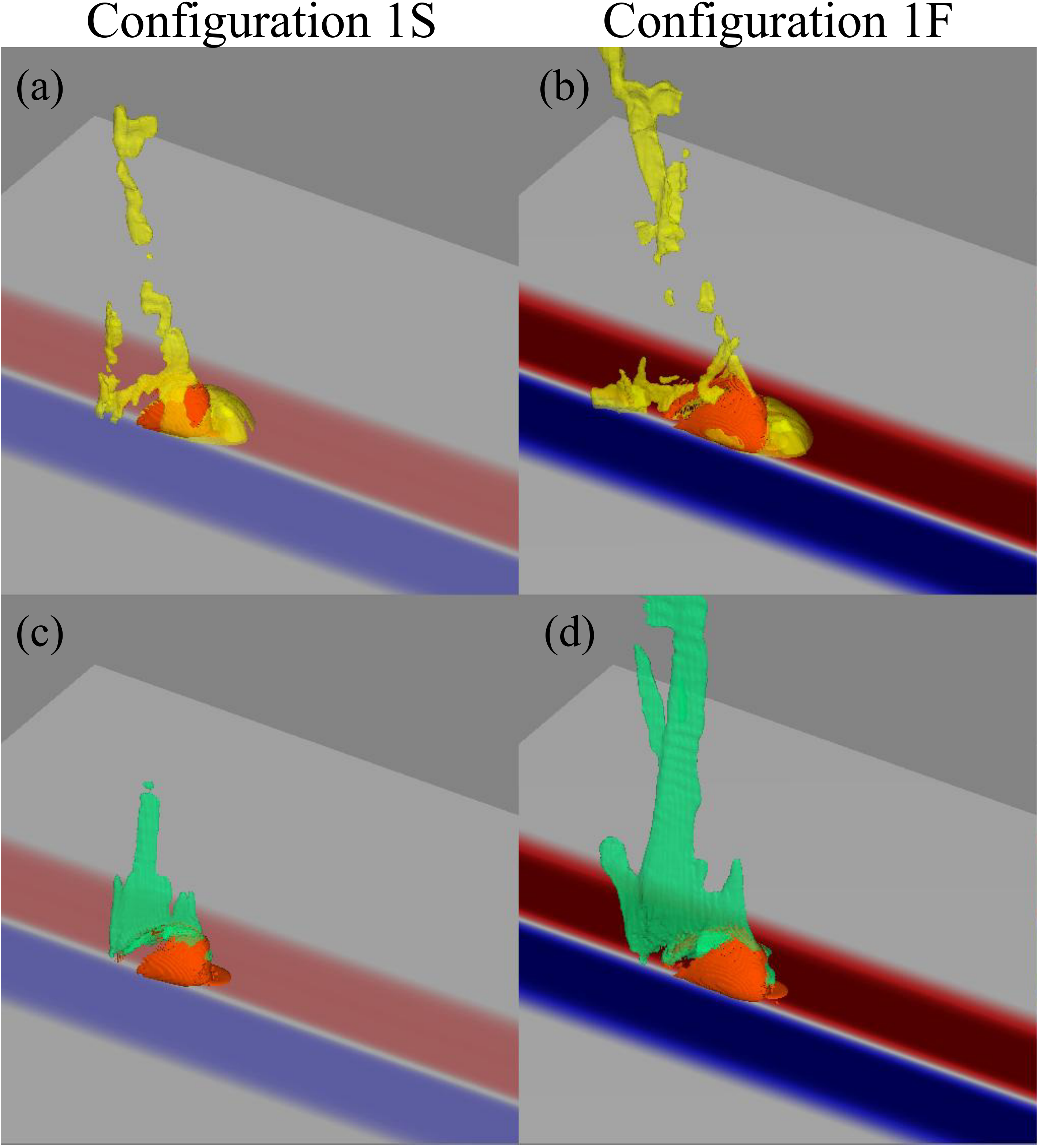}
\caption{Comparison of the jetting/outflows at $t = 360$ in the two simulations for Configuration 1. Top panels: enhanced mass density ($\rho = 1.1$, yellow isosurfaces). Bottom panels: plasma velocity ($v = 0.05$, green isosurfaces). Orange isosurfaces show current density ($J = 0.3$). Red/blue shading shows the surface plasma velocity $v_z$, colour scale as in Figure \ref{fig:initial}(b).}
\label{fig:comp}
\end{figure}

\begin{figure*}
\centering
\includegraphics[width=\textwidth]{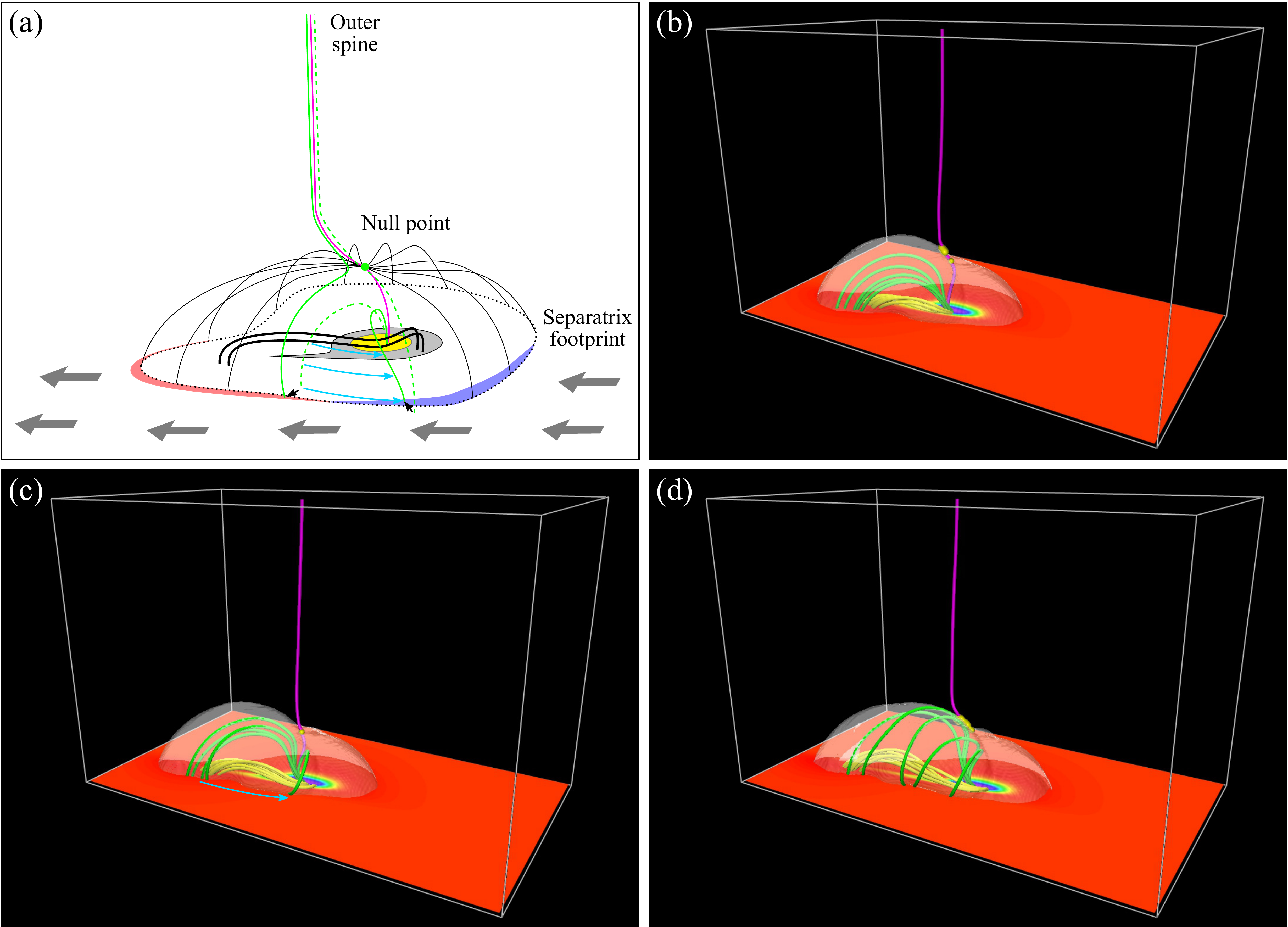}
\caption{(a) Schematic of the initial filament-channel formation process in Configuration 2. See text for details. (b)-(d) Formation of the first filament channel ($t$ = 180, 210, and 270, respectively). Filament-channel field lines are shown in yellow, strapping field lines in green (traced from footpoints in the minority polarity), and field lines near the spine(s) of the null(s) in magenta. Yellow spheres denote the null positions and the semi-transparent isosurface shows the separatrix. Surface shading of $B_x$ is the same as in Figure \ref{fig:simcomp}.}
\label{fig:schematic2}
\end{figure*}

\begin{figure*}
\centering
\includegraphics[width=\textwidth]{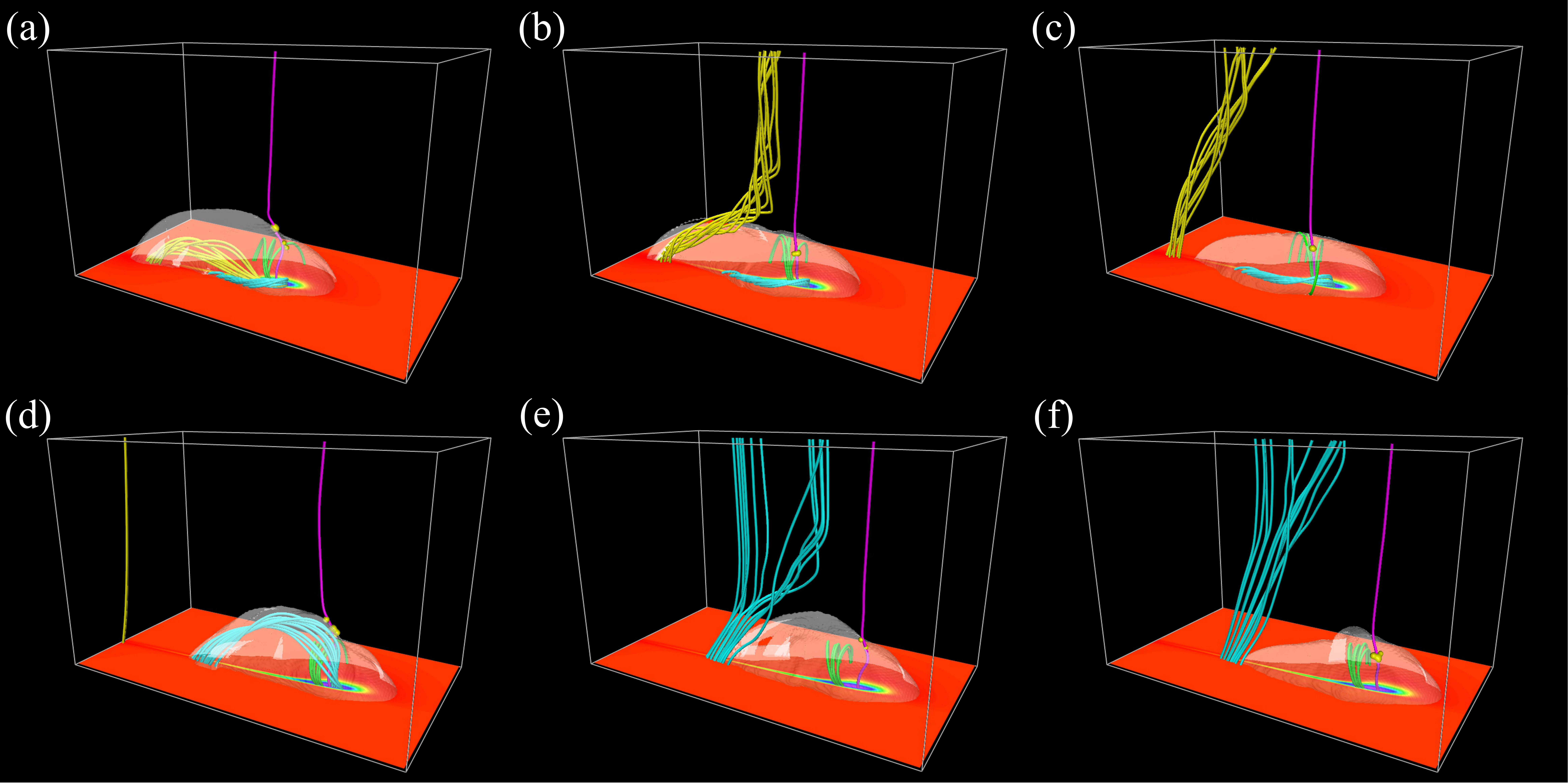}
\caption{Two mini-filament jet eruptions in Configuration 2. (a)-(c) First eruption ($t$ = 340, 360, and 410). (d)-(f) Second eruption ($t$ = 550, 570, and 590). The mini-filaments are drawn with yellow and light blue field lines, respectively. Shading, isosurfaces, and all other field lines as in Figure \ref{fig:schematic2}. An animation is available online.}
\label{fig:fils}
\end{figure*}

Thus, our results indicate that when minority polarity intrusions move relative to a weak background field such that the separatrix and some of the {closed} flux, but not the PIL, are sheared, the interchange reconnection at the null point is characterised by quasi-periodic weak bursts. Each burst arises from a cyclic closing and opening of flux as the minority polarity traverses the background field. We expect such quasi-steady reconnection to continually heat the closed loops along the driven side of the separatrix. Observationally, this should produce a bright-point signature, while the periodic bursts might be observed as periodic increases in intensity (see \S \ref{sec:discussion}). 

In the Configuration 1 simulation, the system adapted to accommodate the rate at which flux is delivered to/removed from the reconnection region by the surface flows. Therefore, the cyclic reconnection dynamics and the assumed intensity modulation may be sensitive to the speed of the moving minority polarity, and also to the efficiency with which the reconnection region can process the arriving magnetic flux. To understand the nature of this relationship, we repeated the experiment with twice the surface driving speed. The driving in this case began to ramp down at $t = 600$, stopping at $t = 650$, so that the minority polarity was advected the same distance. 

Figure \ref{fig:bodyplots} shows the same quantities for this Configuration, 1F, as in Figure \ref{fig:slowplots} for 1S. After a ramp-up phase ($t \le 125$), three cycles of reconnection occurred (seen most easily in peak current density), with the third continuing beyond the end of the driving period. Compared with Configuration 1S, the increased driving speed and reconnection rate of Configuration 1F yielded faster and more extended outflows (Fig. \ref{fig:comp}). 

As occurred with the slow driving, the cycles varied in length, with the middle being slightly longer and exhibiting more than one peak in $v_{max}$. However, the 1F cycles were shorter by a factor of $1.5$ to $2$, similar to the difference in driving speed. This is the key result: for quasi-static driving, {\it i.e.,} slow compared to the coronal Alfven speed, the period of the reconnection cycles is set predominantly by the displacement of the minority polarity, and not by an inherent time scale for the reconnection process. Again this result is similar to what has been found in studies of homologous jets. A burst of reconnection and energy release occurs only after a sufficiently large amount of free energy has built up to trigger some type of feedback between the reconnection and the resulting global dynamics. Since actual photospheric driving velocities are a much smaller fraction of the coronal Alfv\'{e}n speed than those adopted in our simulations, we expect that the cycle period is determined mainly by the flow speed on the Sun, as well.  In our simulations, the minority polarity was displaced by a distance $D \approx 12$, approximately three times the length of the minority polarity patch ($d \approx 4$). Thus, a full cycle occurs every time the minority polarity is advected roughly its length across the solar surface. This is consistent with the picture presented in Figures \ref{fig:simcomp} and \ref{fig:simcomp2} of flux opening along the side of the separatrix in a burst.

\begin{figure*}
\centering
\includegraphics[width=\textwidth]{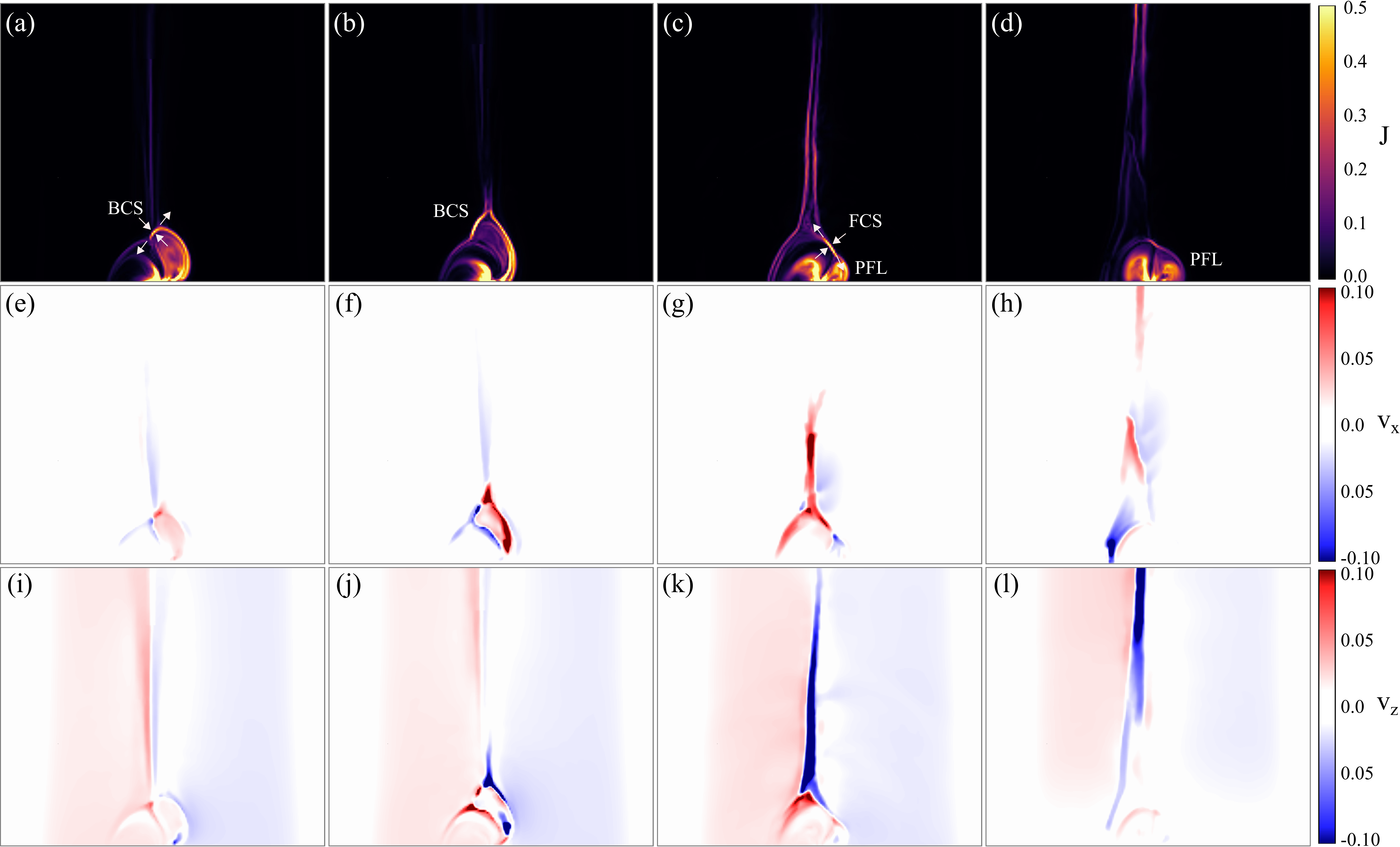}
\caption{Current density $J$ (top), $v_{x}$ (middle), and $v_{z}$ (bottom) in the $z = 1.92$ plane during the second jet in Configuration 2. The panels are ordered from left to right corresponding to $t$ = 490, 520, 580, and 610. Note that the minority polarity moves into the plane as viewed here. BCS = breakout current sheet; FCS = flare current sheet; PFL = post-flare loops. An animation is available online.}
\label{fig:jet}
\end{figure*}

\subsection{Configuration 2 --  Strong Shear}
\label{sec:shear}
In Configuration 2, the flow field was highly nonuniform across the minority-polarity concentration, so that closed field near the PIL was strongly sheared. This is a crucial difference from Configuration 1. The cyclic reconnection described above affects only flux somewhat near the separatrix and the inner-spine footpoint; it does not reach the low-lying field near the PIL. Therefore, in this simulation the stress adjacent to the PIL continued to accumulate until a sheared-arcade filament channel was formed, as illustrated by Figure \ref{fig:schematic2}(a). The yellow patch shows the region of surface flux that is affected by the cyclic reconnection occurring at the null. Green field lines with footpoints in this region undergo interchange reconnection, during which some of their shear is transferred to an open field line and propagates away. The {cyan} arrows show the reduction in shear in one such interchange reconnection event, where the two dashed field lines reconnect to form the two solid field lines. In contrast, the thick black field lines with footpoints near the PIL and outside the yellow region stretch to form a sheared arcade. The weakly sheared, repeatedly reconnecting green field lines form an overlying strapping field that constrains the arcade. Figure \ref{fig:schematic2}(b)-(d) shows the filament channel forming in this manner in the simulation; the filament-channel field lines are yellow, whereas the overlying strapping field lines are green. This leads to the intriguing result that, despite the relatively broad surface shear applied in the simulation, the system naturally forms a low-lying, highly sheared filament channel localised to the PIL, because quasi-steady interchange reconnection at the null continually relaxes the strapping field above. 

Eventually, however, the sheared field of the filament channel builds up sufficient free energy to drive an explosive CME-like release. We identified two explosive eruptions in Configuration 2 before the driving was halted. Figure \ref{fig:fils} shows field lines in the erupting filaments in each case. The first filament (formed as described above) started out as a sheared arcade (Fig.\ \ref{fig:fils}(a)), but as it began to rise it was converted into a flux rope by closed-closed (tether-cutting) reconnection near the PIL. A jet was launched when the portion of the flux rope rooted in the majority polarity opened up by interchange reconnection at the current sheet surrounding the null, releasing the twist in this section as an Alfv\'{e}n wave (Fig.\ \ref{fig:fils}(b)-(c)). The portion of the flux rope rooted in the minority polarity remained closed but shifted to new, less sheared footpoints in the closed region, in a manner similar to the green field lines in Figure \ref{fig:schematic2}(a). This closed section formed the basis of the second filament (light blue field lines), which inherited the remaining twist/shear. Like the first filament channel, this second channel grew in length as the driving continued until eventually it too erupted, generating the second jet. 

\begin{figure*}
\centering
\includegraphics[width=\textwidth]{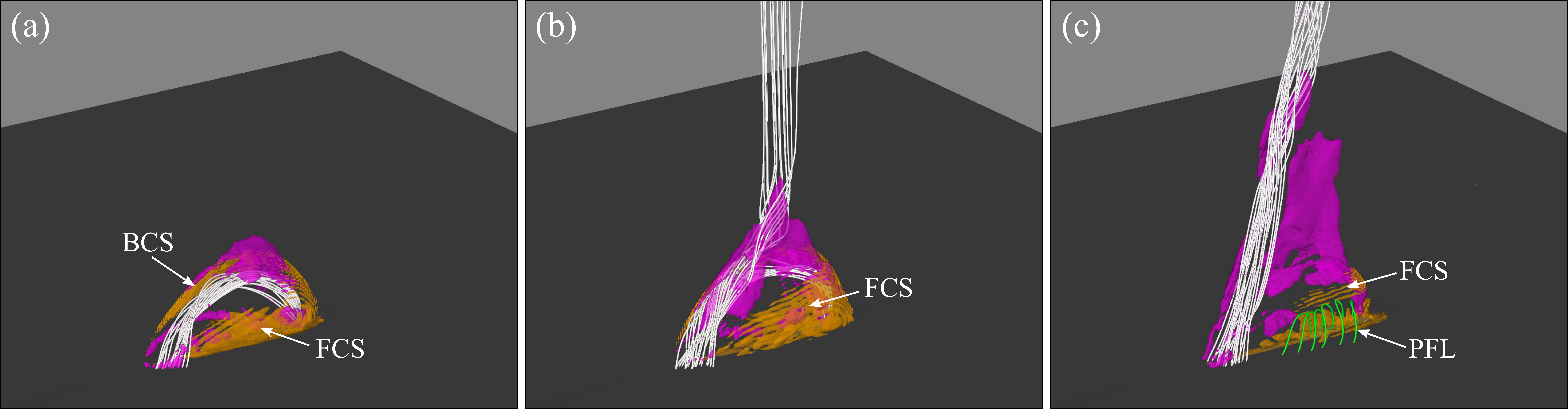}
\caption{3D rendering of the second jet eruption in Configuration 3 at $t$ = (a) 550, (b) 570, and (c) 600. Silver field lines depict the erupting filament. Isosurfaces show plasma velocity ($v = 0.15$, purple) and current density ($J = 0.6$, orange). BCS = breakout current sheet; FCS = flare current sheet; PFL = post-flare loops.}
\label{fig:3d}
\end{figure*}

These eruptions proceeded in the same way as the breakout jets introduced by \citet{Wyper2017,Wyper2018}. Figure \ref{fig:jet}(a)-(d) shows the current density in a plane that cuts across the second jet as the minority polarity moves, depicting the different phases of evolution. The plane is perpendicular to the section of the PIL along which the filament channel is aligned. Just prior to eruption onset (Fig. \ref{fig:jet}(a)), the high volumetric current in the filament channel was visible on the right side of the separatrix dome (outlined by regions of medium strength current). By this stage, the strapping field mostly had been removed from above the filament by reconnection at the null, transferring flux to the unsheared side of the minority polarity and the open field, as shown by the white arrows. Thus, the null current layer acted as a breakout current sheet above the strapping field of the filament, releasing weak reconnection outflows (Fig. \ref{fig:jet}(e) and (i)). The feedback between the rising filament and the reconnection at the null intensified the reconnection outflows as the filament rose (Fig. \ref{fig:jet}(f) and (j)). Finally, a jet was launched when the filament reconnected across the breakout current layer (Fig. \ref{fig:jet}(c)).  At the same time, a flare current layer formed in the wake of the flux rope above growing post-flare loops (Fig. \ref{fig:jet}(c) and (d)).  Interestingly, the jet in this case does not exhibit much untwisting motion (Fig.\ \ref{fig:jet}(l)), instead forming a narrow spire. Both straight and helical jets have been observed, with the latter more frequently associated with mini-filament eruptions \citep{ Moore2010, Sterling2015}. We show a three-dimensional rendering of the current layers and plasma flows in our jet in Figure \ref{fig:3d}.

\begin{figure}
\centering
\includegraphics[width=0.5\textwidth]{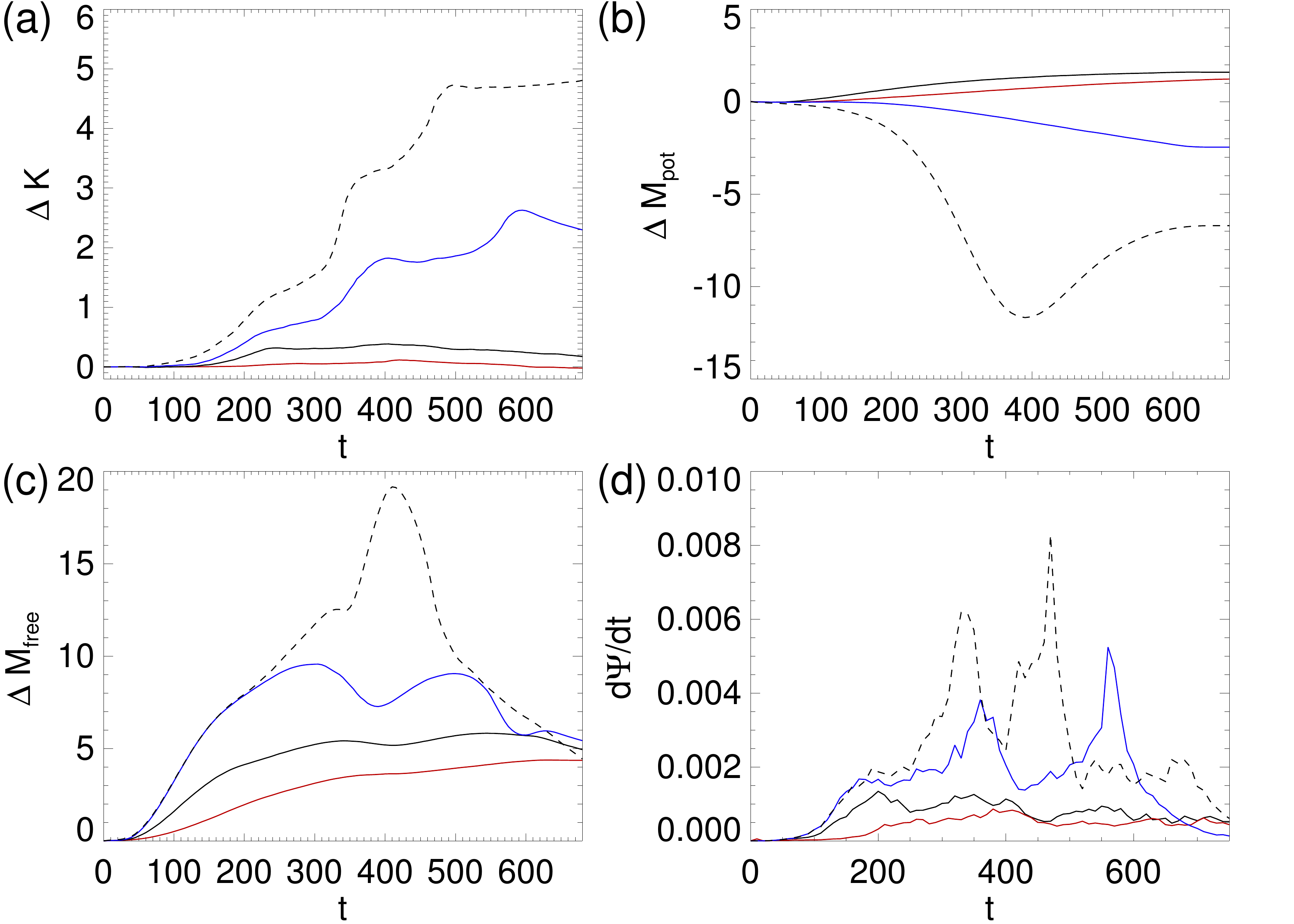}
\caption{Simulation diagnostics. (a) Approximate kinetic energy of the outflows/jets. (b) Change in magnetic potential energy. (c) Approximate free magnetic energy. (d) Normalised reconnection rate. Solid red and black lines: Configurations 1S and 1F, respectively. Solid blue: Configuration 2. Dashed black: Configuration 3.}
\label{fig:energies}
\end{figure}

The jet signatures are prominent in the evolution of free magnetic ($\Delta M_{free}$) and kinetic ($\Delta K$) energy, shown in Figure \ref{fig:energies}, which also shows the change in potential energy ($\Delta M_{pot}$) and the interchange reconnection rate ($d\Psi/dt$). For comparison, the same quantities for the other configurations are included. The nearly uniform advection applied to Configuration 1 generated negligible outflow kinetic energy, whilst the free magnetic energy plateaued at a nearly constant value during each reconnection cycle. The two step increases in kinetic energy and dips in free magnetic energy (starting around $t=330$ and $t=550$) correspond to the onset of the two jets in Configuration 2, as magnetic energy stored in the filaments was released via explosive interchange reconnection (Fig\ \ref{fig:energies}(d)) and converted to collimated plasma motion. Compared to the breakout jets studied by \citet{Wyper2017,Wyper2018}, in which about $50\%$ of the stored magnetic energy was released during the jet, the fractional releases of magnetic energy (and the corresponding increases of kinetic energy) in these jets were significantly smaller. The main difference is that the filament was much smaller in the present simulations than in our previous simulations, straddling only a part of the PIL around the minority polarity rather than wrapping almost entirely around it. Smaller filaments store less free energy, so the resulting jets are less energetic, have narrower spires, and less observable rotation. 

As with our earlier breakout-jet simulations, some magnetic shear remained behind within the filament channel after the jet was launched. This shear provided the foundation of the next filament channel, which formed as the surface motions continued to stress the closed-field region, leading to multiple eruptions. Indeed, these eruptions occurred on their own cycle, similar to the cyclic reconnection observed in Configuration 1. During each breakout jet, the null point rotates around the separatrix from the breakout current layer to the flare current layer as the filament-channel field lines reconnect \citep{Wyper2018}. As the system begins to relax via flare reconnection and the post-flare loops build up, the null (along with its associated spine footpoint) then moves back to its initial position. {We have confirmed this behaviour in Configuration 2 by tracking the evolution of the null centroid (not shown).} In this sense, each jet is simply an extension of the cyclic reconnection scenario of Configuration 1, but with additional reconnection within the closed-field region to form and eject the flux rope.

This simulation demonstrates that, when the field near the PIL of the minority polarity is strongly sheared, complex structure is created inside the separatrix through a combination of interchange (open/closed) reconnection at the null and tether-cutting (closed/closed) reconnection near the PIL. We expect both the overlying loops and the plasma within the filament channel to be heated. This might explain the more complex structure of some bright points observed in predominantly unipolar, open-field regions  \citep[e.g.,][]{Brown2001}. Additionally, our simulation provides a link between jets and bright points, showing how the eruption of the filament channel explosively increases the interchange-reconnection rate and generates a jet via the breakout mechanism.

\begin{figure}
\centering
\includegraphics[width=0.47\textwidth]{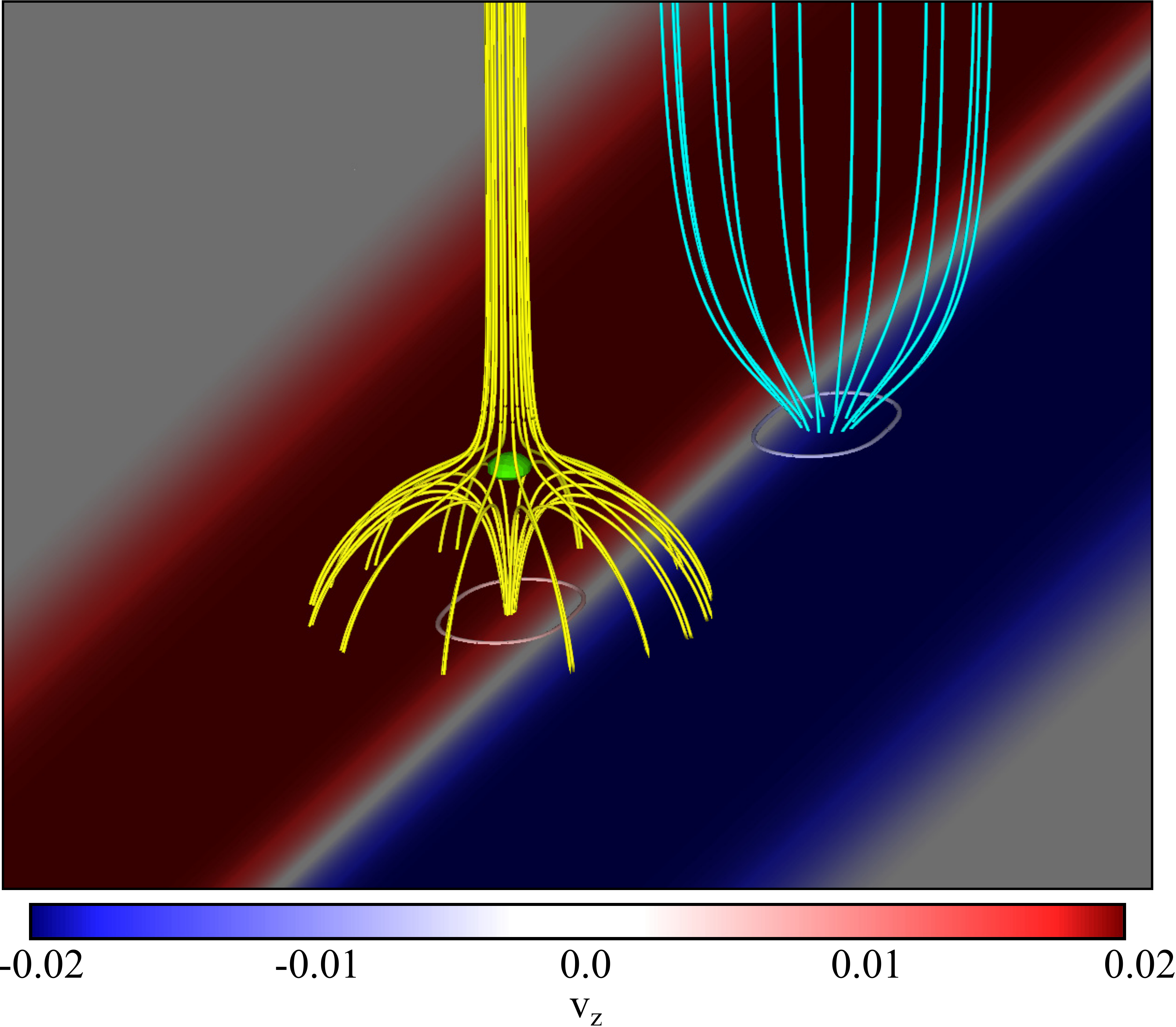}
\caption{The magnetic field and surface flow profile in Configuration 3. Yellow field lines, surface shading and isosurface as in Fig. \ref{fig:initial}. Cyan field lines show the field curvature above the strong majority concentration of surface flux. The PIL is shown as a light pink contour. The light blue contour shows $B_{x} = -2.0$.} 
\label{fig:initial2}
\end{figure}

\begin{figure*}
\centering
\includegraphics[width=\textwidth]{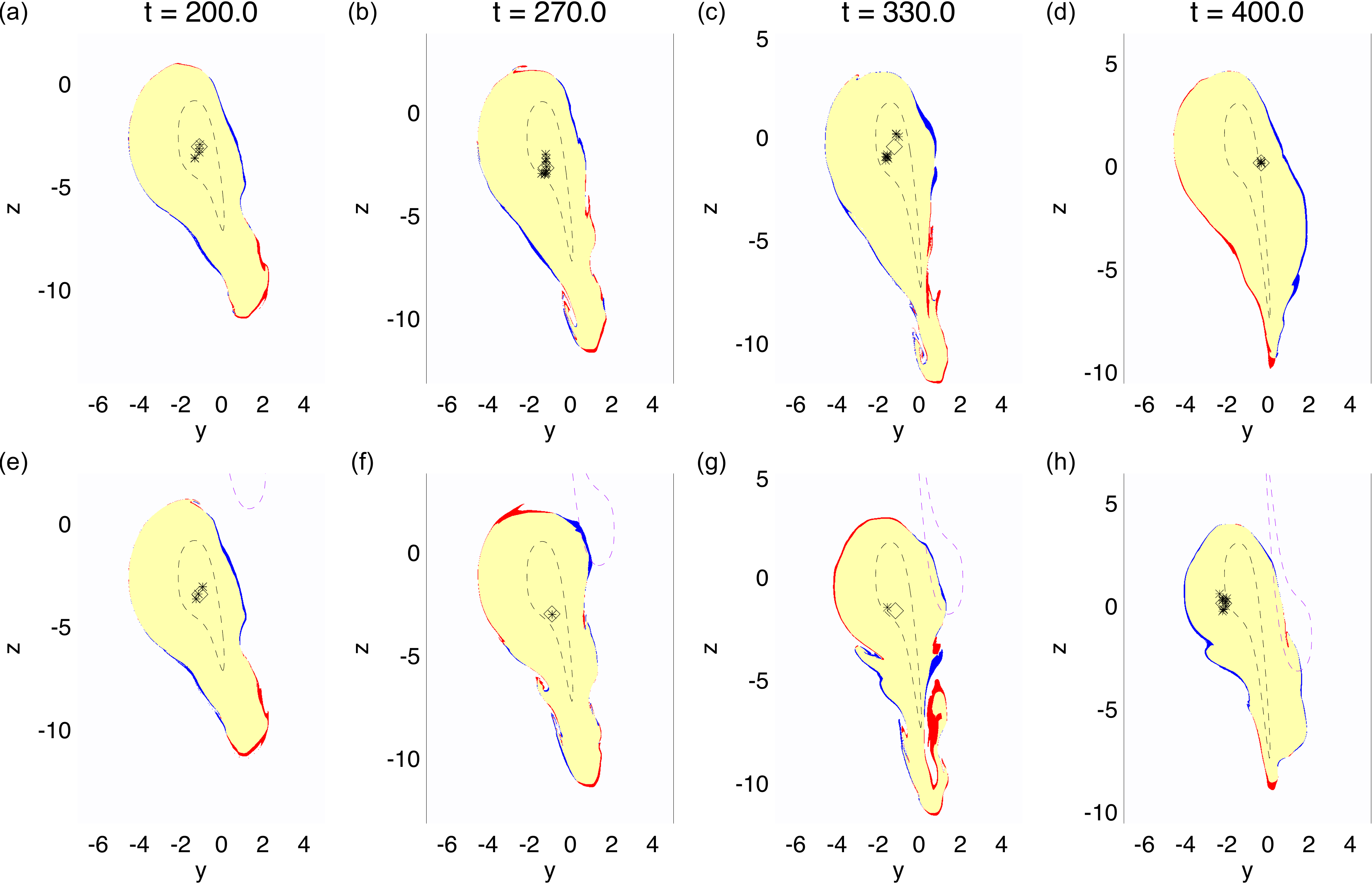}
\caption{The changing flux domains in Configurations 2 (top panels) and 3 (bottom panels) at the same times. Yellow and white show closed and open field regions, respectively. Red and blue show regions of recently {opened and closed field}, respectively. Asterisks denote the projection of the null point positions. The diamond shows the null centroid (smoothed in time). Dashed black lines show the PIL. The dashed purple line in Configuration 3 is a contour of $B_{x} = -2.0$, outlining the majority-polarity (negative) concentration. An animation is available online.}
\label{fig:topology}
\end{figure*}

\begin{figure}
\centering
\includegraphics[width=0.47\textwidth]{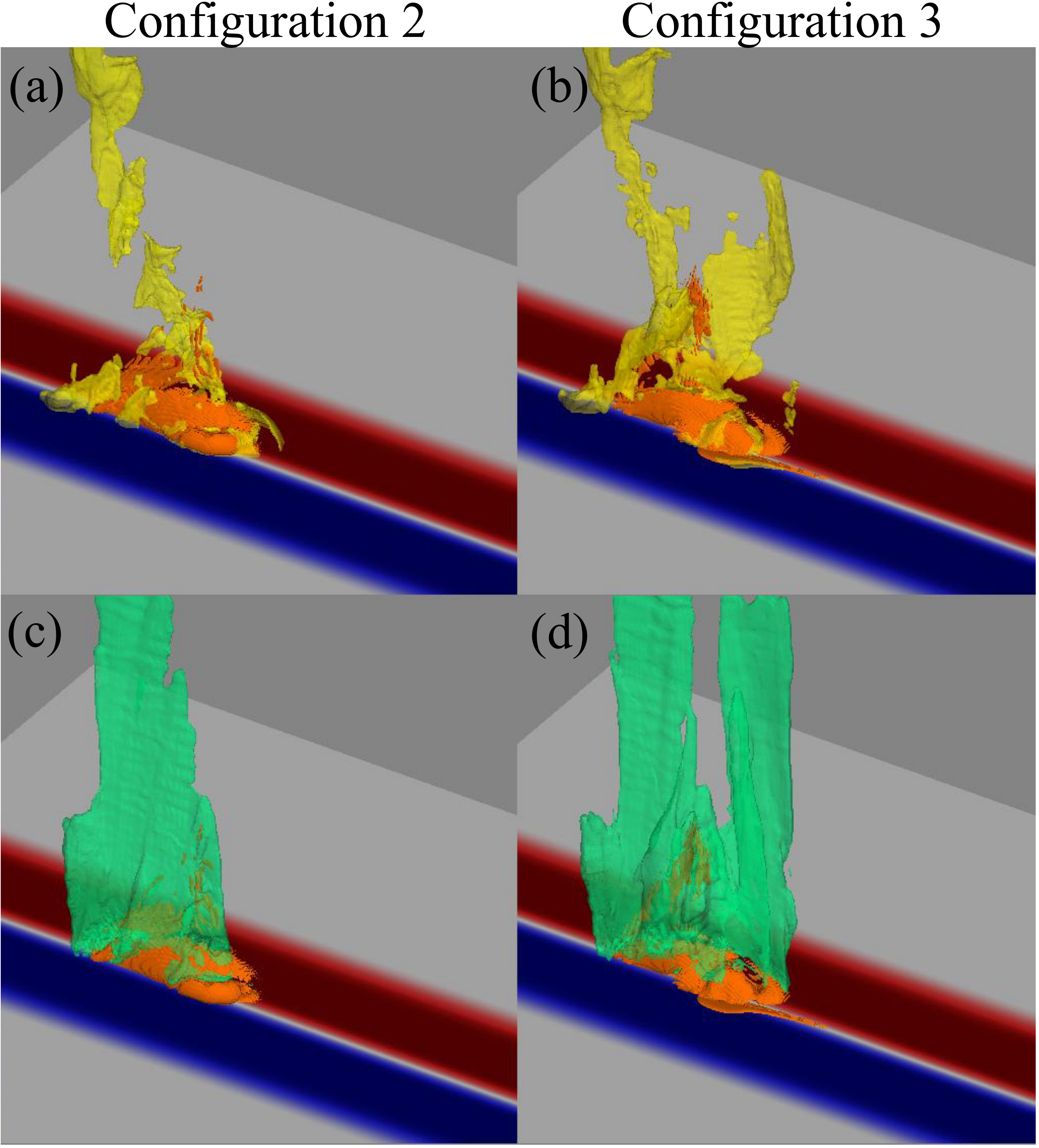}
\caption{Comparison of the jetting/outflows at $t = 360$ in Configurations 2 and 3. Top panels: enhanced mass density ($\rho = 1.1$, yellow isosurfaces). Bottom panels: plasma velocity ($v = 0.05$, green isosurfaces). Orange isosurfaces show current density ($J = 0.3$). Red/blue shading shows the surface plasma velocity $v_z$, colour scale as in Figure \ref{fig:initial}(b).}
\label{fig:comp2}
\end{figure}

\subsection{Configuration 3 -- Fly-by}
\label{sec:fly-by}
In Configurations 1 and 2, we assumed a perfectly uniform majority-polarity background field. However, the actual photospheric field is always clumped into locally strong concentrations. Increased activity is frequently observed when two or more such concentrations interact. We examined such encounters in Configuration 3, by revising Configuration 2 to include a strong concentration of majority-polarity flux. The majority intrusion was placed symmetrically with respect to the minority intrusion  (Fig. \ref{fig:initial2}). As the two polarity concentrations were transported and sheared, they performed a fly-by: approaching, passing, and then receding from each other during the encounter. Because the field above the majority polarity is more concentrated than the background, it expands laterally with height above the surface. Intuitively, we expect that the strong minority- and majority-polarity fields will preferentially establish transient connections as they pass each other. Such connections would be rife in the minimum-energy potential field, when the concentrated minority and majority polarities are closest. On the other hand, as the fly-by finishes and the polarities recede from each other, we expect those connections to be undone. This preferential connection on approach, and disconnection on recession, is a central feature of the bright-point model of \citet{Longcope1998}.

Figure \ref{fig:topology} compares the changing connectivity of the surface flux as the two concentrations passed one another ({bottom} panels) with the changing connectivity of Configuration 2  ({top} panels). Before the majority concentration approached the minority, the flux connectivity was the same in the two cases (Fig. \ref{fig:topology}(a) and (e)). As the majority polarity came closer the connectivities began to differ, with connections forming quickly between the two flux concentrations in Configuration 3. Figures \ref{fig:topology}(f) and (g) show that the background majority-polarity flux opened (red) along the top left of the separatrix when the flux of the majority-polarity concentration closed (blue) along the top right. This burst of interchange reconnection generated a {second weak} outflow at the front of the moving minority polarity,  accompanying the ongoing quasi-steady weak outflows from the rear of the separatrix (Fig. \ref{fig:comp2}(d); compare with Configuration 2, panel \ref{fig:comp2}(c)). The preferential connection to the flux of the passing concentration also shifted the null point away from the majority polarity as the fly-by continued (Fig. \ref{fig:topology}(h); see also the online animation). This shift in the negative $y$ direction changed the null position more than the cyclic motion induced by {the jets}. 

Figure \ref{fig:energies} shows the changes in energy during the fly-by. Most notably, compared with the previous configurations, the potential energy of the system (dashed black line in Fig.\ \ref{fig:energies}(b)) changed substantially, reaching a magnitude similar to the free energy injected by the driving (i.e., roughly equal to $\Delta M_{free}$ in Configuration 2, blue line in Fig.\ \ref{fig:energies}(c)). $\Delta M_{pot}$ dropped to its lowest value as the two polarity patches reached their closest approach, before increasing and finally saturating. This change reflects the connections formed between the two polarity regions, as if the opposing field components became superimposed and nullified each other within the volume as the two concentrations were brought closer together. In this sense, the start of the fly-by is similar in some ways to a cancellation event in which the surface and volumetric fields are both cancelled. The final $\Delta M_{pot}$ is negative because the sheared tails of each polarity remained close to each other after the driving finished (Fig. \ref{fig:bx}). 

The reduction in the potential energy of the system led to substantially more free magnetic energy (Fig. \ref{fig:energies}(c)). This free energy first became available when the two concentrations were close enough to interact (around $t=200$), beyond which the curves of $\Delta M_{free}$ diverge for Configurations 2 and 3. The increased free magnetic energy drove more explosive, energetic jets in Configuration 3. As in Configuration 2, two main breakout jets were produced by bursts of reconnection (Fig.\ \ref{fig:energies}(d)), causing stepped increases in kinetic energy (Fig.\ \ref{fig:energies}(a)). Because the changing potential energy dominates the free magnetic energy in Figure \ref{fig:energies}(c), the drop in $\Delta M_{free}$ during each jet manifests as a pause in the increase of $\Delta M_{free}$ starting around $t = 330$ and a further steepening of the reduction in $\Delta M_{free}$ (arising as the two patches move away from each other) starting around $t = 430$. A third small jet also occurred around $t = 680$, beyond the range of the energy plots but visible in the animation accompanying Figure \ref{fig:topology}. By comparison to Configuration 2, the two main jets occurred in more rapid succession with higher reconnection rates and kinetic energy. However, unlike $\Delta M_{free}$ and $d\Phi/dt$, the volumetric kinetic energy ($\Delta K$) is higher than Configuration 2 even before $t=200$ when the two polarities started to interact. The early increase follows from horizontal plasma flows induced around the base of the majority polarity, which were absent from the reference simulations used to construct the energies and resulted from field lines diverted around flux bulging laterally at the edge of the majority-polarity concentration. These flows were induced by the surface shear in this region, and exhibited no significant jet-like signatures. Once the majority polarity became involved in the jet dynamics, the jet outflows dominated the behaviour of $\Delta K$.

Our results demonstrate that the introduction of a second magnetic-flux concentration, with the same polarity as the background, induces additional reconnection and plasma outflows as the two elements approach and recede from each other. However, the basic feature of a localised filament channel repeatedly forming and erupting via the breakout mechanism, as identified in Configuration 2, remains the same.

\begin{figure*}
\centering
\includegraphics[width=\textwidth]{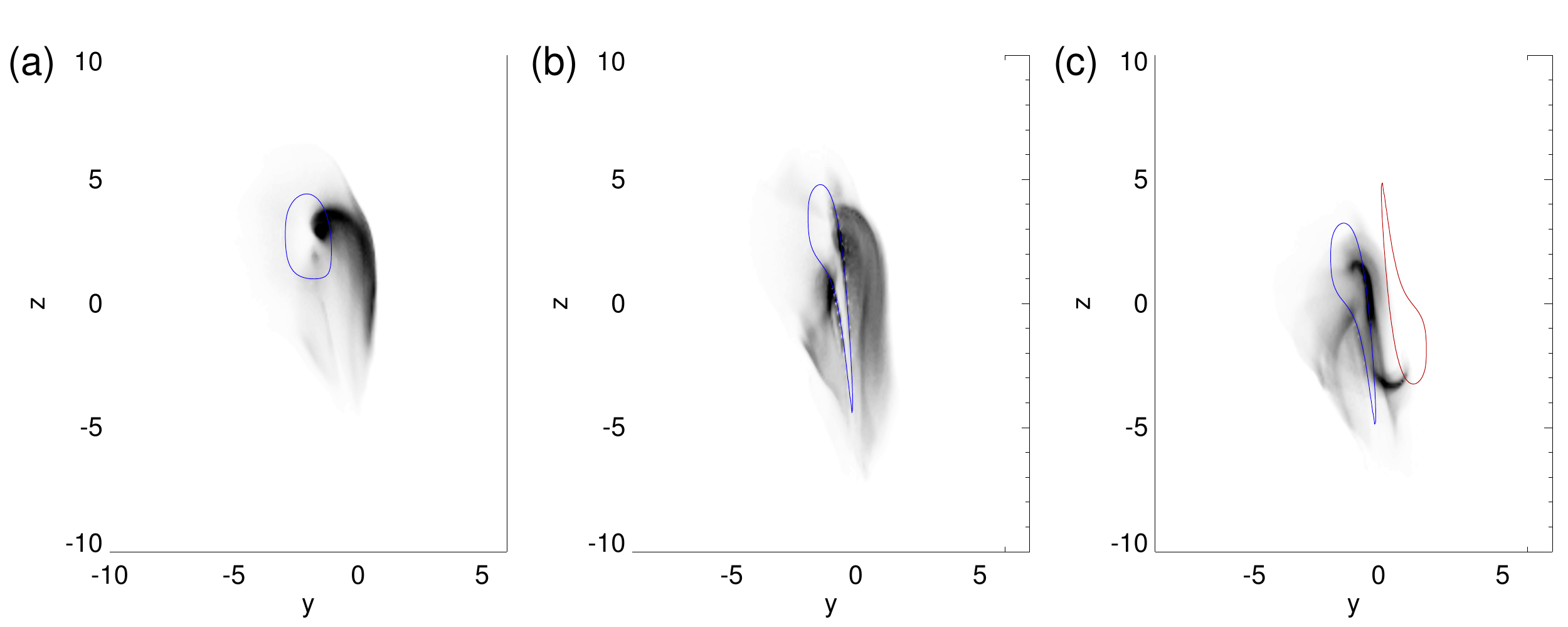}
\caption{Line-of-sight integrated emission proxy for each configuration. (a) Configuration 1S, $t = 900$. (b) Configuration 2, $t = 500$. (c) Configuration 3, $t = 420$. Reverse grey-scale shading shows the emission proxy, scaled to $75\%$ of the maximum in each. Blue and red contours show $B_{x} = 2$ and $-4$, respectively.}
\label{fig:los}
\end{figure*}

\section{Comparison with Observations}
\label{sec:discussion}
We conducted several numerical experiments to understand better the nature of coronal bright points and jets driven by moving magnetic elements. Our simulations differ from most previous bright-point models { \citep[e.g.,][]{Priest1994,Parnell1994,Longcope1998,Galsgaard2000} in that we consider an open ambient field. We can compare our simulations quantitatively to observations by adopting typical values for length scale, field strength, and plasma density as described in \S \ref{sec:setup}. Taking $L_{s}$ = $2.5 \times 10^8$\,cm, $B_{s}$ = 2.5\,G, and $\rho_{s}$ = $4 \times 10^{-16}$\,g cm$^{-3}$ yields $V_{s}$ = $1250$\,km s$^{-1}$, $t_{s}$ = 2\,s, and $E_{s}$ = $9.8 \times 10^{25}$ erg. The width of the separatrix then becomes {$\approx 7 \times L_{s} = 17.5$} Mm and the null initially sits at a height of {$\approx 2.7\times L_{s} = 6.75$ Mm}. Similar values were found by \citet{Zhang2012} in potential field extrapolations above two bright points. The background plasma temperature becomes $T \approx 0.94$ MK and the driving speed translates to $25$ km\,s$^{-1}$ and $12.5$ km\,s$^{-1}$ for the fast and slow speeds, respectively. The minority polarity (now with a peak field strength of $62.5$ G and magnetic flux of {$\approx 8.7\times 10^{18}$\,Mx}) is moved a distance of $12 \times L_{s} = 30$\,Mm in each simulation, comparable to the diameter of a supergranule. }

In Configuration 1S the field close to the separatrix was sheared, producing steady interchange reconnection modulated by quasi-periodic reconnection bursts. {We can roughly estimate the free energy release rate of the steady component from the energy injected before the onset of reconnection {in the first $200$ time units} of the simulation. By $t=200$ around $2$ units of free magnetic energy are injected into the closed field, Fig. \ref{fig:energies}(a). Accounting for the ramp up of the driver and scaling the values this corresponds to an energy injection rate of $\approx 5.6\times 10^{23}$\,erg\,s$^{-1}$ at the maximum driving speed. During the quasi-steady phase this injection is balanced by losses to numerical diffusion, and equates to roughly the free energy available for heating the plasma. Even after accounting for the unrealistically fast driving speed (see below), this energy release rate compares well with the observed values of $10^{23}-10^{24}$\,erg\,s$^{-1}$ for bright points \citep{Golub1974,Priest1994}.} The energy released by the bursts was a small fraction of the stored free magnetic energy -- $\approx 0.5 \times E_{s} = 4.9\times 10^{25}$ erg occurring {with a period of} $\approx 240 \times t_{s} = 8$ min -- whilst the outflow speeds reached typical values of $\approx 0.05 \times V_{s} \approx 60$ km\,s$^{-1}$ along the outer spine. {The energy released in each burst corresponds to $\approx 18\%$ of the energy released over the same period by the steady component.} Many bright points exhibit quasi-periodic intensity increases, with periods ranging from a few minutes to a couple of hours \citep{Kariyappa2008,Tian2008,Zhang2012}. Our results demonstrate that some of this periodicity can be explained by the natural modulation of the interchange reconnection that occurs as minority-polarity elements are moved by surface motions. The predicted outflow speeds, and certainly the periods of the reconnection cycles, are likely too fast because the driving speed ($12.5$ km\,s$^{-1}$) employed in our simulations is too high. However, Configuration 1F demonstrated that the cycle period is mainly set by the displacement of the minority polarity. We speculate that, at more typical photospheric speeds \citep[$\approx 1.5$ km\,s$^{-1}$, e.g.,][]{Brandt1988}, the reconnection cycle period would increase by a factor of $12.5/1.5 \times 8$\,min $\approx 67$\,min, corresponding to the longest observed oscillations in brightness. Without a full treatment of the thermodynamics, however, it is not clear whether the repetitive, low-intensity reconnection jets in this case would be observable.

In Configurations 2 and 3 we showed that filament channels periodically formed and erupted as jets when the field near the PIL is strongly sheared. The jets liberated $\approx 3 \times E_{s} = 2.9 \times 10^{26}$ erg of free magnetic energy, had speeds of $\approx 0.15\times Vs \approx 190$ km\,s$^{-1}$ and durations of $\approx 90\times t_{s} = 2.5$ min. These energies, velocities, and lifetimes are at the lower end of the ranges observed for coronal-hole jets \citep{Shibata1992,Savcheva2007}. Note, however, that our model would predict larger, more energetic jets with different choices for $B_{s}$ and $L_{s}$. Although only two main jets were produced in each simulation, we  expect that further shearing and deposition of energy would continue the cycle. The time between the two jets was comparable to (Configuration 2 $\approx 200\times t_{s}$) or faster than (Configuration 3 $\approx 100\times t_{s}$) the time between the reconnection bursts when only the separatrix was strongly sheared. {However, this was using the faster ($25$\,km\,s$^{-1}$) surface driving speed.} Similarly, we then expect that the time between jets would be roughly a factor of {$25/1.5$ greater, $70$-$140$ min}, for typical solar surface flow speeds. Periodic jets associated with bright point flashes with similar periods were described in \citet{Zhang2012}.

{The periodic bursty dynamics in our simulations follow from the repeated release of energy stored in the closed-field region. Before each burst or jet, the free magnetic energy is built up through an interplay between ideal surface motions shearing the closed field (storing energy) and reconnection opening sheared and closing unsheared field lines (releasing energy). By analysing simulations at different driving speeds we found that the timescale for this storage and release is set primarily by the distance the structures are driven, rather than the driving speed itself. This is true provided that the driving motions are slow compared to the coronal Alfv\'{e}n speed, as occurs on the Sun. Generally speaking, each burst occurs once a threshold of energy storage is reached. The threshold itself is particular to the setup being considered. Once beyond the threshold, some of the energy stored in the closed field is released impulsively over a short time compared with the time for the energy to be stored. The rapid release of energy points to a strong coupling between ideal and non-ideal effects, producing the sharp increase in reconnection rates measured during the bursts and jets. Therefore, ideal and non-ideal effects are present in {\it{both}} the energy buildup phase and the energy release phase. However, it is their coupling in the release phase that leads to the rapid energy releases and bursty dynamics we observe. Such coupling is a general feature of explosive energy release in the corona. It is manifest in these simulations by the upward expansion and ultimate explosive reconnection of the filament channel in the case of jets (configurations 2 and 3), and the similar expansion of the folded field lines and burst of reconnection as they are reconnected in configuration 1. Once a burst or jet occurs each system returns to slowly rebuilding the stored energy via the interplay of storage via ideal shearing and release via relatively slow reconnection around the null.}

In our simulations, the loops of recently reconnected field lines beneath the separatrix are expected to form relatively long-lived bright-point structures. Because our simulations used a simple treatment for the plasma thermodynamics, we cannot directly synthesise observables. However, we obtained a rough estimate of the expected bright-point emission structure in each configuration by using the proxy introduced by \citet{Cheung2012}, whereby the square of the current is averaged along field lines before being integrated along the line of sight to create the image. This procedure picks out current-carrying coronal loops and gives a reasonable comparison to observations in EUV, for example.  In Configuration 1, the relatively unstructured bright point was formed by the recently reconnected loops along one side of the minority polarity (Fig. \ref{fig:los}(a)). In Configuration 2, the main filament channel was localised to the right PIL, producing a bright point with more complex internal structure than Configuration 1 (Fig. \ref{fig:los}(b)). An additional filament channel appeared to the left of the trailing minority-polarity tail at this time, just before the onset of the second eruption. In Configuration 3, a clear sigmoid structure formed between the first and second jet when the filament channel was squeezed between the two passing magnetic elements (Fig. \ref{fig:los}(c)). The uneven magnetic pressure from the element nearest each end of the filament channel distorted the currents into a sigmoid \citep[e.g.,][]{DeVore2000}. Therefore, we have shown that different coronal bright-point morphologies can be realised by altering the way in which the separatrix is sheared and the strength of the flux that passes the minority polarity. The localisation of the bright points to one region beneath the separatrix in some events \citep{Galsgaard2017} may be explained by the results of Configuration 1. On the other hand, bright points with more complex structure and sigmoids \citep[e.g.,][]{Brown2001,Zhang2012} are more consistent with Configurations 2 and 3.

Our modelling also reveals a natural link between bright points and energetic jets. Surprisingly, even when the shear profile was relatively broad, interchange reconnection at the null stripped the shear/helicity from the outer closed field lines to form a localised filament channel adjacent to the PIL. This mechanism is {similar to} the helicity condensation mechanism \citep{Antiochos2013}, whereby volume-filling reconnections drive magnetic shear/twist towards the boundaries where it collects (PILs) or is removed (separatrices). In our case, the boundary (the separatrix) moves in and out to remove the shear adjacent to it. Consequently, the reach of the mechanism is much more limited, ultimately removing most of the helicity through filament-channel eruption and jet formation. During each eruption, the quasi-steady slow reconnection at the null explosively accelerates for a short time, then resumes a slow rate. Interestingly, the jets have little noticeable rotation, which we attribute to the relatively small flux/energy of the filaments compared with the overall flux/energy of the closed-field region. Together with our previous results \citep{Wyper2017,Wyper2018}, this demonstrates that, depending upon the size and energy of the filament channel, both straight and helical jets can be created by the breakout-jet mechanism and thus can be driven by mini-filament eruptions. 

Finally, our results are applicable to other jet- and bright-point-like phenomena involving null points above moving minority-polarity intrusions. At smaller scales, EUV bursts in the chromosphere and transition region exhibit similar features. Recently, \citet{Chitta2017}  identified a moving minority-polarity feature in the moat flow from a sunspot that was sheared by surface flows, and attributed the associated EUV brightening to reconnection driven by the shearing. Our results support their conclusions. At yet smaller scales, continually moving and cancelling minority magnetic elements are associated with tiny jetlets and plume transient bright points observed at the base of plumes \citep{Raouafi2014}. \citet{Raouafi2014} suggested that the collective action of this energy release helps to power and sustain the plumes. Our results might explain the origin of these plume transient bright points and jetlets. Additionally, the compressive and Alfv\'{e}n waves launched by the periodic outflows and homologous jets in our simulations may account for the quasi-periodic waves observed within plumes \citep[e.g.,][]{DeForest1998,Ofman1999,Thurgood2014}.

\section{Summary}
\label{sec:summary}
In this work, we studied minority-polarity moving magnetic elements in an open background field as a model for coronal bright points and jets. Our main results are as follows:
\begin{itemize}
\item All our simulations exhibited the evolution generic to models with the embedded-bipole topology: free energy build up due to ideal stressing by photospheric motions followed by energy release by reconnection at the null and separatrices. The ideal stressing is always very slow, but the reconnection dynamics vary greatly depending on where the free energy builds up, in particular, how close to the separatrix. For example, if the stressing occurs such that only flux very near the separatrix becomes stressed, then the reconnection becomes essentially steady \citep[e.g.][]{Edmondson2010}. On the other hand, if the stress is concentrated far from the separatrix near the PIL, then the evolution must become explosive in order to release this free energy \citep[e.g.,][]{Wyper2017}.    
\item Different bright-point morphologies, from simple loops to sigmoids, can be realised by a combination of the surface shear pattern and the strength and distribution of the flux passing the minority polarity. 
\item Steady interchange reconnection driven by the surface motions is modulated by quasi-periodic, low-intensity reconnection bursts that we speculate would correspond to a quasi-periodic brightening of the newly reconnected bright-point loops. Each burst occurs after the minority polarity has been advected roughly its length across the surface.
\item If the surface motions strongly shear the field near the PIL, a filament channel forms. The bright point produces a jet when the filament channel erupts via the breakout mechanism before returning to long-duration, lower intensity bright point energy release. 
\item Additional bursts of reconnection are driven when strong concentrations of the majority polarity pass by the minority polarity, connecting to and then disconnecting from it. 
\end{itemize}
Our results explain several key aspects of observed coronal bright points and jets, and how the two are related. The results aid in further disentangling the complex behaviour of such events, which also might contribute to the formation and maintenance of coronal plumes. Many potential extensions of this work should be considered, for instance, the role of the background field inclination angle and the implications of including a more realistic treatment of the atmosphere and coronal energy transport processes. Such extensions are left to future work.

\acknowledgments
This work was supported through a Fellowship award to PFW by the Royal Astronomical Society and grant awards to CRD, JTK, and SKA by NASA's Living With a Star and Heliophysics Supporting Research programs. Computer resources for the numerical simulations were provided to CRD by NASA's High-End Computing program at the NASA Center for Climate Simulation. Several of the figures were produced using the Vapor visualisation package (www.vapor.ucar.edu). We are grateful to P.\ Kumar, C.\ E.\ DeForest, N.\ E.\ Raouafi, V.\ M.\ Uritsky, and M.\ A.\ Roberts for numerous helpful discussions regarding jets and plumes. 


\begin{thebibliography}{}
\expandafter\ifx\csname natexlab\endcsname\relax\def\natexlab#1{#1}\fi

\bibitem[{{Antiochos}(1990)}]{Antiochos1990}
{Antiochos}, S.~K. 1990, \memsai, 61, 369

\bibitem[{{Antiochos}(1996)}]{Antiochos1996}
{Antiochos}, S.~K. 1996, in Astronomical Society of the Pacific Conference
  Series, Vol.~95, Solar Drivers of the Interplanetary and Terrestrial
  Disturbances, ed. K.~S. {Balasubramaniam}, S.~L. {Keil}, \& R.~N. {Smartt}, 1

\bibitem[{{Antiochos}(1998)}]{Antiochos1998}
{Antiochos}, S.~K. 1998, \apjl, 502, L181

\bibitem[{{Antiochos}(2013)}]{Antiochos2013}
{Antiochos}, S.~K. 2013, \apj, 772, 72

\bibitem[{{Antiochos} {et~al.}(1999){Antiochos}, {DeVore}, \&
  {Klimchuk}}]{Antiochos1999}
{Antiochos}, S.~K., {DeVore}, C.~R., \& {Klimchuk}, J.~A. 1999, \apj, 510, 485

\bibitem[{{Brandt} {et~al.}(1988){Brandt}, {Scharmer}, {Ferguson}, {Shine}, \&
  {Tarbell}}]{Brandt1988}
{Brandt}, P.~N., {Scharmer}, G.~B., {Ferguson}, S., {Shine}, R.~A., \&
  {Tarbell}, T.~D. 1988, \natur, 335, 238

\bibitem[{{Brown} {et~al.}(2001){Brown}, {Parnell}, {Deluca}, {Golub}, \&
  {McMullen}}]{Brown2001}
{Brown}, D.~S., {Parnell}, C.~E., {Deluca}, E.~E., {Golub}, L., \& {McMullen},
  R.~A. 2001, \solphys, 201, 305

\bibitem[{{Cheung} \& {DeRosa}(2012)}]{Cheung2012}
{Cheung}, M.~C.~M., \& {DeRosa}, M.~L. 2012, \apj, 757, 147

\bibitem[{{Chitta} {et~al.}(2017){Chitta}, {Peter}, {Young}, \&
  {Huang}}]{Chitta2017}
{Chitta}, L.~P., {Peter}, H., {Young}, P.~R., \& {Huang}, Y.-M. 2017, \aap,
  605, A49

\bibitem[{{Craig} \& {McClymont}(1991)}]{Craig1991}
{Craig}, I.~J.~D., \& {McClymont}, A.~N. 1991, \apjl, 371, L41

\bibitem[{{DeForest} \& {Gurman}(1998)}]{DeForest1998}
{DeForest}, C.~E., \& {Gurman}, J.~B. 1998, \apjl, 501, L217

\bibitem[{{DeVore}(1991)}]{DeVore1991}
{DeVore}, C.~R. 1991, \jcoph, 92, 142

\bibitem[{{DeVore} \& {Antiochos}(2000)}]{DeVore2000}
{DeVore}, C.~R., \& {Antiochos}, S.~K. 2000, \apj, 539, 954

\bibitem[{{DeVore} \& {Antiochos}(2008)}]{DeVore2008}
{DeVore}, C.~R., \& {Antiochos}, S.~K. 2008, \apj, 680, 740

\bibitem[{{Doschek} {et~al.}(2010){Doschek}, {Landi}, {Warren}, \&
  {Harra}}]{Doschek2010}
{Doschek}, G.~A., {Landi}, E., {Warren}, H.~P., \& {Harra}, L.~K. 2010, \apj,
  710, 1806

\bibitem[{{Edmondson} {et~al.}(2010){Edmondson}, {Antiochos}, {DeVore}, \&
  {Zurbuchen}}]{Edmondson2010}
{Edmondson}, J.~K., {Antiochos}, S.~K., {DeVore}, C.~R., \& {Zurbuchen}, T.~H.
  2010, \apj, 718, 72

\bibitem[{{Galsgaard} {et~al.}(2017){Galsgaard}, {Madjarska},
  {Moreno-Insertis}, {Huang}, \& {Wiegelmann}}]{Galsgaard2017}
{Galsgaard}, K., {Madjarska}, M.~S., {Moreno-Insertis}, F., {Huang}, Z., \&
  {Wiegelmann}, T. 2017, \aap, 606, A46

\bibitem[{{Galsgaard} {et~al.}(2000){Galsgaard}, {Parnell}, \&
  {Blaizot}}]{Galsgaard2000}
{Galsgaard}, K., {Parnell}, C.~E., \& {Blaizot}, J. 2000, \aap, 362, 395

\bibitem[{{Golub} {et~al.}(1974){Golub}, {Krieger}, {Silk}, {Timothy}, \&
  {Vaiana}}]{Golub1974}
{Golub}, L., {Krieger}, A.~S., {Silk}, J.~K., {Timothy}, A.~F., \& {Vaiana},
  G.~S. 1974, \apjl, 189, L93

\bibitem[{{Habbal} {et~al.}(1990){Habbal}, {Withbroe}, \& {Dowdy}}]{Habbal1990}
{Habbal}, S.~R., {Withbroe}, G.~L., \& {Dowdy}, Jr., J.~F. 1990, \apj, 352, 333

\bibitem[{{Haynes} \& {Parnell}(2007)}]{Haynes2007}
{Haynes}, A.~L., \& {Parnell}, C.~E. 2007, \phpl, 14, 082107

\bibitem[{{Hong} {et~al.}(2014){Hong}, {Jiang}, {Yang}, {Bi}, {Li}, {Yang}, \&
  {Yang}}]{Hong2014}
{Hong}, J., {Jiang}, Y., {Yang}, J., {et~al.} 2014, \apj, 796, 73

\bibitem[{{Hong} {et~al.}(2016){Hong}, {Jiang}, {Yang}, {Yang}, {Xu}, \&
  {Xiang}}]{Hong2016}
{Hong}, J., {Jiang}, Y., {Yang}, J., {et~al.} 2016, \apj, 830, 60

\bibitem[{{Innes} {et~al.}(2009){Innes}, {Genetelli}, {Attie}, \&
  {Potts}}]{Innes2009}
{Innes}, D.~E., {Genetelli}, A., {Attie}, R., \& {Potts}, H.~E. 2009, \aap,
  495, 319

\bibitem[{{Kariyappa} \& {Varghese}(2008)}]{Kariyappa2008}
{Kariyappa}, R., \& {Varghese}, B.~A. 2008, \aap, 485, 289


\bibitem[{{Karpen} {et~al.}(1996){Karpen}, {Antiochos}, \&
  {DeVore}}]{Karpen1996}
{Karpen}, J.~T., {Antiochos}, S.~K., \& {DeVore}, C.~R. 1996, \apjl, 460, L73

\bibitem[{{Karpen} {et~al.}(2012){Karpen}, {Antiochos}, \&
  {DeVore}}]{Karpen2012}
{Karpen}, J.~T., {Antiochos}, S.~K., \& {DeVore}, C.~R. 2012, \apj, 760, 81

\bibitem[{{Karpen} {et~al.}(2017){Karpen}, {DeVore}, {Antiochos}, \&
  {Pariat}}]{Karpen2017}
{Karpen}, J.~T., {DeVore}, C.~R., {Antiochos}, S.~K., \& {Pariat}, E. 2017,
  \apj, 834, 62
  
\bibitem[{{Kumar} {et~al.}(2018){Kumar}, {Karpen}, {Antiochos}, {Wyper},
 {DeVore}, \& {DeForest}}]{Kumar2018}
{Kumar}, P., {Karpen}, J.~T., {Antiochos}, S.~K., {Wyper}, P.~F., {DeVore}, C.~R., 
 \& {DeForest}, C. E. 2018, \apj, 854, 155

\bibitem[{{Lau} \& {Finn}(1990)}]{Lau1990}
{Lau}, Y.~T., \& {Finn}, J.~M. 1990, \apj, 350, 672

\bibitem[{{Longcope}(1998)}]{Longcope1998}
{Longcope}, D.~W. 1998, \apj, 507, 433


\bibitem[{{MacNeice} {et~al.}(2000){MacNeice}, {Olson}, {Mobarry}, {de
  Fainchtein}, \& {Packer}}]{MacNeice2000}
{MacNeice}, P., {Olson}, K.~M., {Mobarry}, C., {de Fainchtein}, R., \&
  {Packer}, C. 2000, \cophc, 126, 330

\bibitem[{{Masson} {et~al.}(2012){Masson}, {Aulanier}, {Pariat}, \&
  {Klein}}]{Masson2012}
{Masson}, S., {Aulanier}, G., {Pariat}, E., \& {Klein}, K.-L. 2012, \soph, 276,
  199

\bibitem[{{Moore} {et~al.}(2010){Moore}, {Cirtain}, {Sterling}, \&
  {Falconer}}]{Moore2010}
{Moore}, R.~L., {Cirtain}, J.~W., {Sterling}, A.~C., \& {Falconer}, D.~A. 2010,
  \apj, 720, 757

\bibitem[{{Mou} {et~al.}(2016){Mou}, {Huang}, {Xia}, {Madjarska}, {Li}, {Fu},
  {Jiao}, \& {Hou}}]{Mou2016}
{Mou}, C., {Huang}, Z., {Xia}, L., {et~al.} 2016, \apj, 818, 9

\bibitem[{{Nistic{\`o}} {et~al.}(2009){Nistic{\`o}}, {Bothmer}, {Patsourakos},
  \& {Zimbardo}}]{Nistico2009}
{Nistic{\`o}}, G., {Bothmer}, V., {Patsourakos}, S., \& {Zimbardo}, G. 2009,
  \solphys, 259, 87

\bibitem[{{Ofman} {et~al.}(1999){Ofman}, {Nakariakov}, \&
  {DeForest}}]{Ofman1999}
{Ofman}, L., {Nakariakov}, V.~M., \& {DeForest}, C.~E. 1999, \apj, 514, 441

\bibitem[{{Pariat} {et~al.}(2009){Pariat}, {Antiochos}, \&
  {DeVore}}]{Pariat2009}
{Pariat}, E., {Antiochos}, S.~K., \& {DeVore}, C.~R. 2009, \apj, 691, 61

\bibitem[{{Pariat} {et~al.}(2010){Pariat}, {Antiochos}, \&
  {DeVore}}]{Pariat2010}
{Pariat}, E., {Antiochos}, S.~K., \& {DeVore}, C.~R. 2010, \apj, 714, 1762

\bibitem[{{Pariat} {et~al.}(2015){Pariat}, {Dalmasse}, {DeVore}, {Antiochos},
  \& {Karpen}}]{Pariat2015}
{Pariat}, E., {Dalmasse}, K., {DeVore}, C.~R., {Antiochos}, S.~K., \& {Karpen},
  J.~T. 2015, \aap, 573, A130

\bibitem[{{Pariat} {et~al.}(2016){Pariat}, {Dalmasse}, {DeVore}, {Antiochos},
  \& {Karpen}}]{Pariat2016}
{Pariat}, E., {Dalmasse}, K., {DeVore}, C.~R., {Antiochos}, S.~K., \& {Karpen},
  J.~T. 2016, \aap, 596, A36

\bibitem[{{Parnell} {et~al.}(1994){Parnell}, {Priest}, \&
  {Golub}}]{Parnell1994}
{Parnell}, C.~E., {Priest}, E.~R., \& {Golub}, L. 1994, \solphys, 151, 57

\bibitem[{{Pontin} {et~al.}(2007){Pontin}, {Bhattacharjee}, \&
  {Galsgaard}}]{Pontin2007}
{Pontin}, D.~I., {Bhattacharjee}, A., \& {Galsgaard}, K. 2007, \phpl, 14, 052106

\bibitem[{{Pontin} {et~al.}(2005){Pontin}, {Hornig}, \& {Priest}}]{Pontin2005}
{Pontin}, D.~I., {Hornig}, G., \& {Priest}, E.~R. 2005, \gapfd, 99, 77

\bibitem[{{Pontin} {et~al.}(2013){Pontin}, {Priest}, \&
  {Galsgaard}}]{Pontin2013}
{Pontin}, D.~I., {Priest}, E.~R., \& {Galsgaard}, K. 2013, \apj, 774, 154

\bibitem[{{Priest} {et~al.}(1994){Priest}, {Parnell}, \& {Martin}}]{Priest1994}
{Priest}, E.~R., {Parnell}, C.~E., \& {Martin}, S.~F. 1994, \apj, 427, 459

\bibitem[{{Pucci} {et~al.}(2012){Pucci}, {Poletto}, {Sterling}, \&
  {Romoli}}]{Pucci2012}
{Pucci}, S., {Poletto}, G., {Sterling}, A.~C., \& {Romoli}, M. 2012, \apjl,
  745, L31

\bibitem[{{Raouafi} {et~al.}(2010){Raouafi}, {Georgoulis}, {Rust}, \&
  {Bernasconi}}]{Raouafi2010}
{Raouafi}, N.-E., {Georgoulis}, M.~K., {Rust}, D.~M., \& {Bernasconi}, P.~N.
  2010, \apj, 718, 981

\bibitem[{{Raouafi} {et~al.}(2008){Raouafi}, {Petrie}, {Norton}, {Henney}, \&
  {Solanki}}]{Raouafi2008}
{Raouafi}, N.-E., {Petrie}, G.~J.~D., {Norton}, A.~A., {Henney}, C.~J., \&
  {Solanki}, S.~K. 2008, \apjl, 682, L137

\bibitem[{{Raouafi} \& {Stenborg}(2014)}]{Raouafi2014}
{Raouafi}, N.-E., \& {Stenborg}, G. 2014, \apj, 787, 118

\bibitem[{{Raouafi} {et~al.}(2016){Raouafi}, {Patsourakos}, {Pariat}, {Young},
  {Sterling}, {Savcheva}, {Shimojo}, {Moreno-Insertis}, {DeVore}, {Archontis},
  {T{\"o}r{\"o}k}, {Mason}, {Curdt}, {Meyer}, {Dalmasse}, \&
  {Matsui}}]{Raouafi2016}
{Raouafi}, N.~E., {Patsourakos}, S., {Pariat}, E., {et~al.} 2016, \ssr, 201, 1

\bibitem[{{Savcheva} {et~al.}(2007){Savcheva}, {Cirtain}, {Deluca},
  {Lundquist}, {Golub}, {Weber}, {Shimojo}, {Shibasaki}, {Sakao}, {Narukage},
  {Tsuneta}, \& {Kano}}]{Savcheva2007}
{Savcheva}, A., {Cirtain}, J., {Deluca}, E.~E., {et~al.} 2007, \pasj, 59, 771

\bibitem[{{Shen} {et~al.}(2012){Shen}, {Liu}, {Su}, \& {Deng}}]{Shen2012}
{Shen}, Y., {Liu}, Y., {Su}, J., \& {Deng}, Y. 2012, \apj, 745, 164

\bibitem[{{Shibata} {et~al.}(1994){Shibata}, {Nitta}, {Strong}, {Matsumoto},
  {Yokoyama}, {Hirayama}, {Hudson}, \& {Ogawara}}]{Shibata1994}
{Shibata}, K., {Nitta}, N., {Strong}, K.~T., {et~al.} 1994, \apjl, 431, L51

\bibitem[{{Shibata} {et~al.}(1992){Shibata}, {Ishido}, {Acton}, {Strong},
  {Hirayama}, {Uchida}, {McAllister}, {Matsumoto}, {Tsuneta}, {Shimizu},
  {Hara}, {Sakurai}, {Ichimoto}, {Nishino}, \& {Ogawara}}]{Shibata1992}
{Shibata}, K., {Ishido}, Y., {Acton}, L.~W., {et~al.} 1992, \pasj, 44, L173

\bibitem[{{Shimojo} {et~al.}(1996){Shimojo}, {Hashimoto}, {Shibata},
  {Hirayama}, {Hudson}, \& {Acton}}]{Shimojo1996}
{Shimojo}, M., {Hashimoto}, S., {Shibata}, K., {et~al.} 1996, \pasj, 48, 123


\bibitem[{{Sterling} {et~al.}(2015){Sterling}, {Moore}, {Falconer}, \&
  {Adams}}]{Sterling2015}
{Sterling}, A.~C., {Moore}, R.~L., {Falconer}, D.~A., \& {Adams}, M. 2015,
  \natur, 523, 437

\bibitem[{{Sterling} {et~al.}(2016){Sterling}, {Moore}, {Falconer}, {Panesar},
  {Akiyama}, {Yashiro}, \& {Gopalswamy}}]{Sterling2016}
{Sterling}, A.~C., {Moore}, R.~L., {Falconer}, D.~A., {et~al.} 2016, \apj, 821,
  100

\bibitem[{{Thurgood} {et~al.}(2014){Thurgood}, {Morton}, \&
  {McLaughlin}}]{Thurgood2014}
{Thurgood}, J.~O., {Morton}, R.~J., \& {McLaughlin}, J.~A. 2014, \apjl, 790, L2

\bibitem[{{Thurgood} {et~al.}(2017){Thurgood}, {Pontin}, \&
  {McLaughlin}}]{Thurgood2017}
{Thurgood}, J.~O., {Pontin}, D.~I., \& {McLaughlin}, J.~A. 2017, \apj, 844, 2

\bibitem[{{Tian} {et~al.}(2008){Tian}, {Xia}, \& {Li}}]{Tian2008}
{Tian}, H., {Xia}, L.-D., \& {Li}, S. 2008, \aap, 489, 741

\bibitem[{{Titov} {et~al.}(2009){Titov}, {Forbes}, {Priest}, {Miki{\'c}}, \&
  {Linker}}]{Titov2009}
{Titov}, V.~S., {Forbes}, T.~G., {Priest}, E.~R., {Miki{\'c}}, Z., \& {Linker},
  J.~A. 2009, \apj, 693, 1029

\bibitem[{{Wang} {et~al.}(1998){Wang}, {Sheeley}, {Socker}, {Howard},
  {Brueckner}, {Michels}, {Moses}, {St.~Cyr}, {Llebaria}, \&
  {Delaboudini{\`e}re}}]{Wang1998}
{Wang}, Y.-M., {Sheeley}, Jr., N.~R., {Socker}, D.~G., {et~al.} 1998, \apj,
  508, 899

\bibitem[{{Webb} {et~al.}(1993){Webb}, {Martin}, {Moses}, \&
  {Harvey}}]{Webb1993}
{Webb}, D.~F., {Martin}, S.~F., {Moses}, D., \& {Harvey}, J.~W. 1993, \solphys,
  144, 15

\bibitem[{{Wyper} {et~al.}(2017){Wyper}, {Antiochos}, \& {DeVore}}]{Wyper2017}
{Wyper}, P.~F., {Antiochos}, S.~K., \& {DeVore}, C.~R. 2017, \natur, 544, 452

\bibitem[{{Wyper} \& {DeVore}(2016)}]{Wyper2016}
{Wyper}, P.~F., \& {DeVore}, C.~R. 2016, \apj, 820, 77

\bibitem[{{Wyper} {et~al.}(2018){Wyper}, {DeVore}, \& {Antiochos}}]{Wyper2018}
{Wyper}, P.~F., {DeVore}, C.~R., \& {Antiochos}, S.~K. 2018, \apj, 852, 98

\bibitem[{{Wyper} {et~al.}(2016){Wyper}, {DeVore}, {Karpen}, \&
  {Lynch}}]{Wyper2016b}
{Wyper}, P.~F., {DeVore}, C.~R., {Karpen}, J.~T., \& {Lynch}, B.~J. 2016, \apj,
  827, 4

\bibitem[{{Wyper} {et~al.}(2012){Wyper}, {Jain}, \& {Pontin}}]{Wyper2012}
{Wyper}, P.~F., {Jain}, R., \& {Pontin}, D.~I. 2012, \aap, 545, A78


\bibitem[{{Young}(2015)}]{Young2015}
{Young}, P.~R. 2015, \apj, 801, 124

\bibitem[{{Zhang} {et~al.}(2001){Zhang}, {Kundu}, \& {White}}]{Zhang2001}
{Zhang}, J., {Kundu}, M.~R., \& {White}, S.~M. 2001, \solphys, 198, 347

\bibitem[{{Zhang} {et~al.}(2012){Zhang}, {Chen}, {Guo}, {Fang}, \&
  {Ding}}]{Zhang2012}
{Zhang}, Q.~M., {Chen}, P.~F., {Guo}, Y., {Fang}, C., \& {Ding}, M.~D. 2012,
  \apj, 746, 19


\end{thebibliography}

\end{document}